\documentclass[aps,prd,twocolumn,floatfix,nofootinbib,groupedaddress,superscriptaddress,a4paper]{revtex4}
\usepackage{amsmath, amssymb, graphics}

\newcommand{\mathsym}[1]{{}}
\newcommand{\unicode}[1]{{}}
\usepackage{textcomp}
\usepackage{eurosym}
\usepackage[utf8]{inputenc}
\usepackage{amsfonts}
\usepackage{array}
\usepackage{amsthm}
\usepackage{bm}
\usepackage{helvet}

\usepackage{graphicx}
\usepackage{subfigure}
\usepackage{amsmath, amsfonts, amssymb}
\usepackage{bm, array}
\usepackage{amssymb}
\usepackage[T1]{fontenc}
\usepackage[colorlinks=true,
            linkcolor=blue,
           urlcolor=blue,
           citecolor=blue]{hyperref}
\def\be{\begin{equation}}
\def\ee{\end{equation}}
\def\beq{\begin{eqnarray}}
\def\eeq{\end{eqnarray}}

\newcommand{\sfrac}[2]{{\textstyle{#1\over #2}}}
\newcommand{\ds}{\displaystyle}
\newcommand{\ben}{\begin{eqnarray}}
\newcommand{\een}{\end{eqnarray}}
\newcommand{\vphi}{\psi}

\def\R{{}^3\!R}
\newtheorem{thm}{Theorem}

\newtheorem{rem}{Remark}
\newtheorem{prop}{Proposition}
\newtheorem{defn}{Definition}

\begin{document}

\title{The past and future dynamics of quintom dark energy models}
\author{Genly Leon}\email{genly.leon@ucn.cl}\affiliation{Departamento de Matem\'aticas, Universidad Cat\'olica del Norte, Avda. Angamos 0610, Casilla 1280 Antofagasta, Chile}
\author{Andronikos Paliathanasis} \email{anpaliat@phys.uoa.gr}
\affiliation{ Instituto de Ciencias F\'{\i}sicas y Matem\'aticas, Universidad Austral de Chile, Valdivia, Chile}
  \affiliation{Department of Mathematics and Natural Sciences, Core Curriculum Program, Prince
Mohammad Bin Fahd University, Al Khobar 31952, Kingdom of Saudi Arabia}
\affiliation{Institute of Systems Science, Durban University of Technology, PO Box 1334, Durban 4000, South
Africa}                    
\author{Jorge Luis Morales-Mart\'inez}\email{jorge.morales@cimat.mx}\affiliation{Dpto. de Geomática e Hidráulica, Divisi\'on de Ingenier\'ia, Universidad de Guanajuato, Guanajuato, M\'exico}
\date{\today}

\begin{abstract}
We study the phase space of the quintom cosmologies for a class of exponential potentials. We combine normal forms expansions and the center manifold theory in order to describe the dynamics near equilibrium sets. Furthermore, we construct the unstable and center manifold of the massless scalar field cosmology motivated by the numerical results given in Lazkoz and Leon (Phys Lett B 638:303. arXiv:astro-ph/0602590, 2006). We study the role of the curvature on the dynamics. Several monotonic functions are defined on relevant invariant sets for the quintom cosmology. Finally, conservation laws of the cosmological field equations and algebraic solutions are determined by using the symmetry analysis and the singularity analysis.
\end{abstract}

\maketitle


\section{Introduction}

Recent observations coming from type I-a Supernovae (SNIa), large scale structure (LSS) formation, and the cosmic microwave background (CMBR) Radiation \cite{1,2,3,4,5,6,7,8,9,10,11,12,13,14,15,16}, indicates the universe is expanding with an accelerated rate without been settled until now a definitive model to account for this phenomena. One of the current alternatives models for such acceleration of the expansion is the yet unknown form of matter named dark energy. Several models of dark energy use a single scalar field, e.g., the so called phantom field and the quintessence field \cite{17,18}. Furthermore, one can consider models with two fields, in particular combining quintessence and phantom fields one can built up the so-called quintom models \cite{19,20,21,22,23,24,25,26,27,28,29,30,31,32,33,34,35,36,37,38,39,40,41,42,43,44,45}. These models admits the crossing of the boundary $w=-1$ with $w$ being the equation of state parameter of the cosmic fluid. This behavior has been investigated in several cosmological theories, for example: h-essence cosmologies\cite{28,29}; holographic dark energy \cite{46,47,48,49}; inspired by string theory \cite{33}; by considering spinor matter \cite{37,38}; for arbitrary potentials \cite{40,41,42,43,44}; using isomorphic models consisting of coupled oscillators \cite{50}; Pais–Uhlenbeck oscillator \cite{51}; inspired in scalar tensor theories \cite{52,53,54,55,56} as well as in modified theories of gravity \cite{57,58}. Furthermore, interacting quintom in nonminimal coupling has been studied in \cite{59}, and quintom with nonminimal derivative coupling was studied in \cite{60}. An extension of the quintom scenario, in which the scalar fields are coupled through a kinetic interaction, have been studied in \cite{61,62} while an interacting quintom model was recently constructed by the application of the generalized uncertainty principle (GUP) in the scalar field Lagrangian [63]. In particular, the second scalar field of the quintom model in \cite{63} corresponds to higher-order derivatives given by the GUP. Furthermore, one of the particularly exciting solutions of quintom type models is that they can provide a classically stable bouncing solution. This was first pointed out in \cite{61} and two important extensions were made in \cite{62} and \cite{63}, which discussed the Lee-Wick and Horndeski realizations, respectively. The cosmology of the quintom model with exponential potentials was investigated in \cite{30} and \cite{31,32}. More specifically, \cite{31,32} the potential accounts for the interaction between the conventional scalar
field and the phantom field. In contrary with \cite{30} where no such interaction was assumed. In \cite{31} was found that the interaction term dominates over the mixed terms of the potential.This simplification has important consequences such that the existence of scaling attractors excludes the existence of phantom attractors, as a difference with the results in \cite{32}. Some of these results were extended in \cite{40}, for arbitrary potentials. In the recent years still there an special interest in this kind of models with the phantom crossing behavior \cite{67,68,69,70,71,72,73}. In line with this interest, we stress in this paper a
deep insight about the model proposed in \cite{31}. 
The plan of the paper it follows. In Section \ref{frame} we make a general description of the model under
study and we present the field equations. In Section \ref{IIB} are presented
several monotonic functions defined on invariant sets of the phase space
using the results summarized in the Appendix \ref{monotonic-th}. The equilibrium points of the model
for any curvature choice using $H$-normalization are discussed in Section %
\ref{IIA}. In Section %
\ref{two} we make the analysis of zero-curvature models with two scalar
fields, one a linear function of the other.  Furthermore, in section \ref{sectionIII} we present some improvements of
the results from reference \cite{31}. In particular, in
Section \ref{normal} we discuss the normal expansion
with undetermined coefficients of arbitrary order that is used to construct the unstable and center manifold up to fourth order in \cite{Leon:2008aq} for flat FRW metric. We
explicitly state in the Appendix \ref{NFTheory} the main techniques to use,
following the approach of \cite{arrowsmith,wiggins,wiggins2}. In section \ref%
{CompactNewflat} we introduce an alternative of the variables used in \cite%
{31} for flat FRW models that renders the phase space in compact
form. In this new variables we find static universe configurations that cannot by discussed in the framework of \cite{31}.  This illustrates the applicability of this formulation in compact variables. Moreover, in Section \ref{crossing} we propose several
normalization procedures for investigating models with flat and non flat
spatial curvature, that allows to circumvent some technical problems that
arises, for example when the crossing through $H=0$, where $H$ is the Hubble
factor, happens.
The integrability of the quintom cosmological
model is studied in Section \ref{integrability}. In particular we apply two different methods
to determine the integrability, the symmetry analysis and the singularity
analysis. Finally, in Section \ref{conclusions} we discuss our results and
draw our conclusions.

\section{The quintom model}

\label{frame}

In this work we consider the so-called two-field quintom with Lagrangian \cite{31}:
\begin{align}
& \mathcal{L}=\frac{1}{2}\partial _{\mu }\phi \partial ^{\mu }\phi -\frac{1}{%
2}\partial _{\mu }\psi \partial ^{\mu }\psi -V(\phi ,\psi ),
\label{expquintomlag} \\
& V(\phi ,\psi )=V_{0}e^{-\sqrt{6}(m\phi +n\psi )},
\label{ExponentialPot}
\end{align}%
where the scalar field $\phi $ represents quintessence and $\psi $
represents a phantom field, and we include a co-moving perfect fluid in the
gravitational action.  We use the Friedmann-Robertson-Walker (FRW) line element given in
spherical coordinates by: {\small
\begin{equation}
ds^{2}=-dt^{2}+a(t)^{2}\left( \frac{\mathrm{d}r^{2}}{1-kr^{2}}+r^{2}\mathrm{d%
}\theta ^{2}+r^{2}\sin ^{2}\theta \mathrm{d}\vartheta ^{2}\right) .
\label{FRW}
\end{equation}%
}
The scalar curvature of the 3-metric induced on the space-like surfaces orthogonal to ${\bf e}_0=\partial_t$ is given by
$\R = \frac{6k}{a^2},\quad
   k = 1,0,-1,$ where $k$ identifies the three types of FRW universes: closed, flat, and open, respectively.

The field equations derived for the line element (\ref{FRW}) in the context
of General Relativity, are given by
\begin{align}
& H^2 - \sfrac16\left(\dot\phi^2-\dot\psi^2\right)-\sfrac13 V(\phi,\psi)-\sfrac13\rho=-\sfrac{\R}{6},\nonumber\\
&\dot H=-H^2-\sfrac13\left(\dot\phi^2-\dot\psi^2\right)+\sfrac13 V(\phi,\psi)-\sfrac16 (3\gamma-2)\rho,\nonumber\\
&\dot\rho=-3\gamma H\rho,\nonumber\\
&\ddot\phi+3H\dot\phi-\sqrt{6}m V(\phi,\psi)=0,\nonumber\\
&\ddot\psi+3H\dot\psi+\sqrt{6}n V(\phi,\psi)=0,\label{system}
\end{align} where $H =\frac{\dot a(t)}{a(t)}$ denotes the Hubble function, while the dot denotes differentiation with respect to variable $t$. Moreover, the barotropic index of the background matter, $\gamma,$ is assumed to take values on the interval $0<\gamma<2.$ 

For the stability analysis of cosmological models one can apply linearization around fixed points, Monotonic Principle \cite{LeBlanc:1994qm}, the Invariant
Manifold Theorem \cite{reza,arrowsmith,wiggins,wiggins2,aulbach}, the Center Manifold Theorem
\cite{arrowsmith,carr,wiggins,wiggins2}, and Normal Forms (NF) theory \cite{arrowsmith,wiggins,wiggins2}. 

Introducing the following normalized variables
\begin{equation}
x_\phi=\frac{\dot\phi}{\sqrt{6}H},\;x_\vphi=\frac{\dot\vphi}{\sqrt{6}H},\; y=\frac{\sqrt V}{\sqrt 3 H},\;z=\frac{\sqrt {\rho}}{\sqrt 3 H},\label{00vars}
\end{equation} the Friedmann equation becomes
\begin{equation}
-1+x_\phi^2- x_\vphi^2+y^2+z^2=\sfrac{\R}{6H^2},
\label{ct}
\end{equation}
and the deceleration factor $q\equiv-\ddot a a/\dot a^2$ can be expressed in terms of the new variables as 
\be q=\sfrac12 (3\gamma-2)z^2+2\left(x_\phi^2-x_\psi^2\right)-y^2.\label{decc}\ee 

The new variables (\ref{00vars}) can be used to rewrite the system (\ref{system}) as a system of first-order ordinary differential equations (ODEs):

\begin{eqnarray}
&&x_\phi'=\sfrac13 \left(3 m y^2+(q-2) x_\phi\right),\label{0eqxphi}\\
&&x_\vphi'=-\sfrac13 \left(3 n y^2-(q-2) x_\psi \right),\label{0eqxvphi}\\
&&y'=\sfrac13 (1+q-3(m x_\phi+n x_\psi)) y, \label{0eqy}\\
&&z'=\sfrac16(1+2 q-3\gamma)z,  \label{0eqz}
\end{eqnarray} where the primes denote derivative with respect to  $\tau=\log a^3.$

\subsection{Invariant set and monotonic functions}
\label{IIB}
By following the procedures in the Appendix \ref{monotonic-th}, we can find the invariant sets. 
\begin{rem} The sets 
$\left\{\mathbf{x}\in\mathbb{R}^4| x_\phi^2- x_\vphi^2+y^2+z^2\lesseqqgtr
1\right\}$, $\left\{\mathbf{x}\in\mathbb{R}^4| n x_\phi+m x_\psi\lesseqqgtr
0\right\}$,  $\left\{\mathbf{x}\in\mathbb{R}^4| y\lesseqqgtr
0\right\}$, and $\left\{\mathbf{x}\in\mathbb{R}^4| z\lesseqqgtr
0\right\}$ are invariant sets of   (\ref{0eqxphi}-\ref{0eqz}).
\end{rem} 
By using the equations 
$\frac{d}{d\tau}\ln (-1+x_\phi^2- x_\vphi^2+y^2+z^2)=\frac{2}{3} q$,
and 
$\frac{d}{d\tau}\ln (n x_\phi+m x_\psi)=\frac{1}{3}(q-2)$, it follows from Proposition \ref{Proposition 4.1} that the sign of $\R$ and the sign of $n x_\phi+m x_\psi$ are invariant. From equations (\ref{0eqxphi}-\ref{0eqz}) it is obvious that any combination of signs for $y$ and $z$ is an invariant set for the flow.
\begin{rem}
For $\R=0$,
$\left\{\mathbf{x}\in\mathbb{R}^3| x_\phi^2- x_\vphi^2+y^2\leqq
1\right\}$ are invariant sets of the flow of  (\ref{0eqxphi}-\ref{0eqz}).
\end{rem}
 Defining on this sets the function $Z=x_\phi^2-x_\psi^2+y^2-1$ that has directional derivative along the flow given by $Z'=\alpha Z$ where $\alpha$ is a continuous real-valued non-vanishing function on $\mathbb{R}^3$ given by $\alpha=x_\phi^2-x_\psi^2-y^2$. Hence, by applying Proposition \ref{Proposition 4.1}, the desired result follows.  In the same way it is proved that $y=0$ is an invariant set of the flow by defining $\alpha=1+q-3\left(mx_\phi+nx_\psi\right)$. 
Combining the former results we can prove that the 2-dimensional set $\Gamma=\left\{\mathbf{x}\in\mathbb{R}^3: 0<x_\phi^2-x_\psi^2<1,y=0\right\}$ is invariant under the flow. Notice that the equilibrium points $O$ and $C_\pm$ studied in \cite{31} are in the boundary of $\Gamma$.

Below: $\Phi({\bf x}_0,t)$ is the unique solution of the system satisfying $\Phi({\bf x}_0,0)={\bf x}_0$;   $O({\bf x}_0)$ denotes the unique orbit passing through ${\bf x}_0$; $\alpha({\bf x}_0)$ means the past attractor or $\alpha$-limit of ${\bf x}_0$, defined by $\alpha({\bf x}_0)=\{{\bf y}: \exists \{t_n\}_{n\geq 1}, t_n\rightarrow -\infty, \lim_{n\rightarrow +\infty} \Phi({\bf x}_0,t_n)={\bf y}\}$; and $\omega({\bf x}_0)$ means the future attractor or $\omega$-limit of ${\bf x}_0$, defined by $\omega({\bf x}_0)=\{{\bf y}: \exists \{t_n\}_{n\geq 1}, t_n\rightarrow +\infty, \lim_{n\rightarrow +\infty} \Phi({\bf x}_0,t_n)={\bf y}\}$.

\begin{prop}
Let ${\bf x}_0\in S_1:=\{{\bf x}: n x_\phi+m x_\psi\neq 0, x_\phi^2- x_\vphi^2+y^2+z^2\neq 1\}$ such that $O({\bf x}_0)$ is bounded. Then, $\alpha({\bf x}_0)\subset \{{\bf x}: x_\phi^2- x_\psi^2+y^2+z^2=1\}$ and $\omega({\bf x}_0)\subset \{{\bf x}:n x_\phi+m x_\psi=0\}.$ 
\end{prop}
{\bf Proof}. Let 
$W_1=\frac{\left(n x_\phi+m x_\psi\right)^4}{\left(x_\phi^2- x_\vphi^2+y^2+z^2-1\right){2}}, W_1'=-\sfrac83 W_1.$ Thus, $W_1$ is positive and strictly monotone along orbits of the invariant set $S_1$. The boundary of $S_1$ is the invariant set $\bar{S_1}\setminus S_1=\{{\bf x}: n x_\phi+m x_\psi= 0\}\cup\{{\bf x}: x_\phi^2- x_\vphi^2+y^2+z^2=1\}$. Let be any initial state ${\bf x}_0\in S_1$ such that $O({\bf x}_0)$ remains bounded. Therefore, $a:=\inf_{{\bf x}\in S_1}W_1({\bf x})=0$ is attained at $\{{\bf x}: n x_\phi+m x_\psi= 0\}$ and $b:=\sup_{{\bf x}\in S_1}W_1({\bf x})=\infty$ is attained at $\{{\bf x}: x_\phi^2- x_\vphi^2+y^2+z^2=1\}$. By the Monotonicity Principle (Theorem \ref{theorem 4.12}), $\alpha({\bf x}_0)\subset \{{\bf x}: x_\phi^2- x_\psi^2+y^2+z^2=1\}$, $\omega({\bf x}_0)\subset \{{\bf x}:n x_\phi+m x_\psi=0\}.$ 
$\blacksquare$

\begin{prop}
Consider ${\bf x}_0\in S_2:=\{{\bf x}: z> 0,\;nx_\phi+mx_\psi\neq 0\}$ such that $O({\bf x}_0)$ is bounded. Then $\alpha({\bf x}_0)\subset \{{\bf x}: z=0\}$  and $\omega({\bf x}_0)\subset \{{\bf x}:n x_\phi+m x_\psi=0\}.$ 
\end{prop}
{\bf Proof}.
Let
$W_2=\frac{\left(n x_\phi+m x_\psi\right)^2}{z^{2}}, W_2'=(\gamma-2) W_2.$ Thus, $W_2$ is positive and strictly monotone along orbits of the invariant set $S_2$. Hence, the boundary of $S_2$ is the invariant set $\bar{S_2}\setminus S_2=\{{\bf x}: n x_\phi+m x_\psi= 0\}\cup\{{\bf x}: z=0\}$. Let be any initial state ${\bf x}_0\in S_1$ such that $O({\bf x}_0)$ remains bounded. By Theorem \ref{theorem 4.12}, $\alpha({\bf x}_0)\subset \{{\bf x}: z=0\}$, $\omega({\bf x}_0)\subset \{{\bf x}:n x_\phi+m x_\psi=0\}.$ $\blacksquare$

\begin{prop}
Let ${\bf x}_0\in S_3:=\{{\bf x}: x_\phi^2- x_\vphi^2+y^2+z^2\neq 1,\; z>0\}$ such that $O({\bf x}_0)$ is bounded. Then:
\begin{itemize}
\item if $\gamma>\sfrac23$, $\alpha({\bf x}_0)\subset \{ {\bf x}: x_\phi^2- x_\vphi^2+y^2+z^2=1\}$ and $\omega({\bf x}_0)\subset \{ {\bf x}:z=0\}$. 

\item if $\gamma<\sfrac23$, $\alpha({\bf x}_0)\subset \{ {\bf x}:z=0\}$ and $\omega({\bf x}_0)\subset \{ {\bf x}: x_\phi^2- x_\vphi^2+y^2+z^2=1\}$.

\end{itemize}
\end{prop}
{\bf Proof}. Let  
$W_3=\frac{z^4}{\left(x_\phi^2- x_\vphi^2+y^2+z^2-1\right)^{2}}, W_3'=-\sfrac23 (3\gamma-2)W_3.$ $W_3$ is positive monotonic decreasing (increasing) in the invariant set $S_3$ for non-zero curvature models provided $\gamma>\sfrac23$ ($\gamma<\sfrac23$). $\bar{S_3}\setminus S_3= \{{\bf x}:z=0\}\cup \{ {\bf x}: x_\phi^2- x_\vphi^2+y^2+z^2=1\}$.  Let be  ${\bf x}_0\in S_3$ such that $O({\bf x}_0)$ is bounded. By Theorem \ref{theorem 4.12} we conclude. $\blacksquare$

\begin{prop}
Let ${\bf x}_0\in S_4:=\left\{(x_\phi,\,x_\psi,y): 0<x_\phi^2-x_\psi^2<1,y=0\right\}$, such that $O({\bf x}_0)$ is bounded. Then, $\alpha({\bf x}_0)\subset C:=\{{\bf x}: x_\phi^2-x_\psi^2=1\}$ and $\omega({\bf x}_0)\subset \{{\bf x}:n x_\phi+m x_\psi=0\}.$
\end{prop}
{\bf Proof.}
Let be $W_4=\frac{(nx_\phi+mx_\psi)^4}{(1-x_\phi^2+x_\psi^2)^2}, W_4'=\frac{4(q-2)}{3(1-x_\phi^2+x_\psi^2)}W_4.$ Notice $q=\sfrac12\left(3(x_\phi^2-x_\psi^2)+1\right)<2$ since $y=0$ and $0<x_\phi^2-x_\psi^2<1.$ Then,  $W_4$ is positive and strictly monotone along orbits of the invariant set $S_4$. $\bar{S_4}\setminus S_4=\{{\bf x}: n x_\phi+m x_\psi= 0\}\cup\{{\bf x}: x_\phi^2- x_\vphi^2=1\}$.  Let be any initial state ${\bf x}_0\in S_4$ such that $O({\bf x}_0)$ remains bounded. By Theorem \ref{theorem 4.12}  we find  $\alpha({\bf x}_0)\subset C:=\{{\bf x}: x_\phi^2-x_\psi^2=1\}$ and $\omega({\bf x}_0)\subset \{{\bf x}:n x_\phi+m x_\psi=0\}.$ $\blacksquare$

\begin{rem}
In \cite{31} it was devised the monotonic function $
W_5=\frac{x_\phi^4}{(1-x_\phi^2+x_\psi^2)^2}, {W_5}'=-2{W_5},$ defined in the invariant set of zero-curvature models with dust matter. There it was argued that the orbits initially near the hyperbolas (representing massless scalar field cosmologies) tends asymptotically to the flat FRW universe.  On the other hand, from the monotonic function $W_4$ we can conclude that (for zero-curvature models) the orbits which initially near the hyperbolas tends asymptotically to either the matter scaling (MS) solutions (equilibrium points like $T$) or (in the absence of scaling solutions) to powerlaw solutions (equilibrium points like $P$). 
\end{rem}

\begin{rem}
The previous gallery of monotonic functions can be used to ruled out, according to Proposition \ref{Proposition 4.2}, the existence of homoclinic orbits, recurrent orbits and more complex behavior in several invariant sets. The dynamics is dominated by equilibrium points  and heteroclinic orbits joining it. Therefore, some global results can be found from the linear (local) analysis of equilibrium points.
\end{rem}


\subsection{The equilibrium points}
\label{IIA}

The system (\ref{0eqxphi}-\ref{0eqz}) admits six types of equilibrium points (with both $z$ and $y$ non-negative) denoted by $M,\;F,\;C_\pm,\;P,\; MS,$ and $CS.$ In tables \ref{0tab1} and \ref{0tab2} we offer some partial information of these equilibrium points like the location, conditions for existence, dynamical character and some additional information. For zero-curvature models (with dust background, i.e., $\gamma=1$), the equilibrium point $MS$ and $P$ reduces respectively to the points $T$ and $P,$ investigated in \cite{31}.  

The equilibrium point $M$ represents the Milne universe, it has negative curvature. It is no hyperbolic if $\gamma=\sfrac{2}{3}.$ $F$ denotes the flat FRW universe and it is no hyperbolic if $\gamma=\sfrac{2}{3}$ or $\gamma=2.$ For $\gamma<2/3$ (resp. $\gamma>2/3$) the eigenvalues of $M$ (resp. $F$) are $(-,-,+,+)$ and the eigenvalues of $F$ (resp. $M$) are $(-,-,+,-)$. In this case both equilibrium points act locally as saddles. The 3-dimensional stable subspace of $F$ (resp. $M$) intersects the 2-dimensional unstable subspace of $M$ (resp. $F$) at the $z$ axis. This means that the orbits initially at this axis are asymptotic to the past to the Milne (resp. flat FRW) universe and to the flat FRW (resp. Milne) universe toward the future. 

The equilibrium sets given by the hyperbolas $C_\pm$ were investigated in \cite{31}, the same analysis done in that reference is valid in our more general context. In the former section we conjecture that they are local sources for the quintom phase-space. In fact, if $\delta>1$ then part of the hyperbola is a local attractor to the past, and part of it is a saddle. If $\delta<1$ then the entire hyperbola is a local attractor to the past.  

The equilibrium point $P$ denotes a zero-curvature power-law inflationary model if $\delta<1/3$, corresponding to the equilibrium point with the same label investigated in \cite{31}.  It is worth noticing that the results of \cite{31} concerning to this equilibrium point applies here only in the case of zero-curvature models with a dust background. It is non hyperbolic if $\delta=\sfrac{1}{3}$ or $\delta=\sfrac{\gamma}{2}.$   $P$ is a local inflationary attractor if either $0<\gamma<2/3,\;\delta<\gamma/2$ or $2/3<\gamma<2,\; \delta<1/3.$ If $2/3<\gamma<2$ and $1/3<\delta<\gamma/2$ then it is no longer a local inflationary solution nor an attractor for non-zero curvature models. However in this case the eigenvalues are $(+,-,-,-)$ and then, the equilibrium point has a 3-dimensional stable subspace. 
If $0<\gamma<2/3$ and $\gamma/2<\delta<1/3$ the eigenvalues of $P$ are $(-,-,-,+).$ Hence, although an inflationary solution, $P$ is not longer a local attractor. In this case the stable subspace is 3-dimensional.

The equilibrium point $MS$ represents a matter scaling model. It is no hyperbolic if $\gamma=0$ or $\gamma={2}/{3}$ or $\gamma=2.$  If $2/3<\gamma<2$ the eigenvalues are $(-,+,-,-).$ In this case the equilibrium point behaves like a saddle, the stable subspace being 3-dimensional. Otherwise the equilibrium point is stable. If either $0<\gamma\leq {2}/{9},\;\delta>{\gamma}/{2}$  or if ${2}/{9}<\gamma\leq {2}/{3},\;{\gamma}/{2}<\delta\leq {4\gamma^2}/{(9\gamma-2)}$, the equilibrium point behaves as an stable node. If ${2}/{9}<\gamma< {2}/{3},\;\delta>{4\gamma^2}/{(9\gamma-2)}$ there are two complex eigenvalues and the orbits spiral-in toward $MS$ in a 2-dimensional subspace.

As stated in \cite{31}, point $P$ is the local attractor for zero-curvature models provided there are not scaling regimes. As we have seen, for some region in the parameters space the equilibrium points $P$ and $MS$ exist in the same phase portrait with $P$ having eigenvalues $(-,-,-,+)$ (i.e, a saddle) and $MS$ being a local attractor (i.e., all its eigenvalues having positive real part). If $0<\gamma\leq 2/3,\; 1/3<\delta<1$ or $2/3<\gamma<2,\; \gamma/2<\delta<1$ the eigenvalues of $P$ are $(+,-,-,+).$ In the latter case, the stable subspace of $P$ is 2 dimensional and then it is the local attractor for zero-curvature models without matter.

Furthermore, the equilibrium point $CS$ denotes curvature scaling models. It is no hyperbolic if $\gamma={2}/{3}$ or $\delta={3}/{7}.$ It is stable if  ${2}/{3}<\gamma<2,\;\delta>{3}/{7},$ and then, with negative curvature.  If ${2}/{3}<\gamma<2,\;{3}/{7}<\delta\leq {4}/{9},$ then the eigenvalues are negative real and $CS$ behaves as a stable node. If ${2}/{3}<\gamma<2,\;\delta>{4}/{9}$ the orbits spiral-in toward $CS$ in a 2-dimensional subspace (since in this case two eigenvalues are complex conjugated). Otherwise, $CS$ is unstable. Its stable subspace is 3-dimensional if either $\gamma>2/3,\; 0<\delta<3/7$ or $0<\gamma<2/3,\; 3/7 <\delta\leq 4/9$. This subspace is 2-dimensional if $0<\gamma<2/3,\; 0<\delta<3/7.$ 
 \subsubsection{A flat FRW model with one scalar field a linear function of the other}\label{two}
Consider the 2-dimensional invariant set $n x_\phi+m x_\psi=0, x_\phi^2-x_\psi^2+y^2+z^2=1$. Let $m\neq 0$ such that $x_\psi=-nx_\phi/m$. Therefore, $\left(1-\frac{n^2}{m^2}\right)x_\phi^2+y^2+z^2=1$. 
The dynamics is governed by the equations:
\begin{align}  x_\phi'&=\left(1-\sfrac{\gamma}{2}\right)x_\phi\left(\left(1-\sfrac{n^2}{m^2}\right) x_\phi^2-1\right) +\left(m-\sfrac{\gamma}{2}x_\phi\right)y^2,\label{eq13}\\
y'&=-\frac{1}{2} y \Big[x_\phi (2 m+x_\phi (\gamma -2))
   \left(1-\sfrac{n^2}{m^2}\right)+\left(y^2-1\right) \gamma \Big],\label{eq14}\end{align} 
defined on the phase plane:
$\Psi=\left\{(x_\phi,y): 0\leq \left(1-\frac{n^2}{m^2}\right)x_\phi^2+y^2\leq 1,\; y\geq 0\right\}.$
We have two cases: Case i) if $m^2\geq n^2,$ the phase plane $\Psi$ is the half of the ellipse $0\leq (1-{n^2}/{m^2}) x_\phi^2+y^2\leq 1$.
Case ii) if $m^2< n^2,$ the phase space is half of the hyperbola $0\leq (1-{n^2}/{m^2}) x_\phi^2+y^2\leq 1.$ 
 The conditions for existence and the dynamical behavior of the equilibrium points are presented in Table \ref{i_table1}.
 
\begin{figure*}[ht]
\centering
\subfigure[\;\;$m=0.6, n=0.5,\gamma=1$.]{\includegraphics[width=0.3\textwidth]{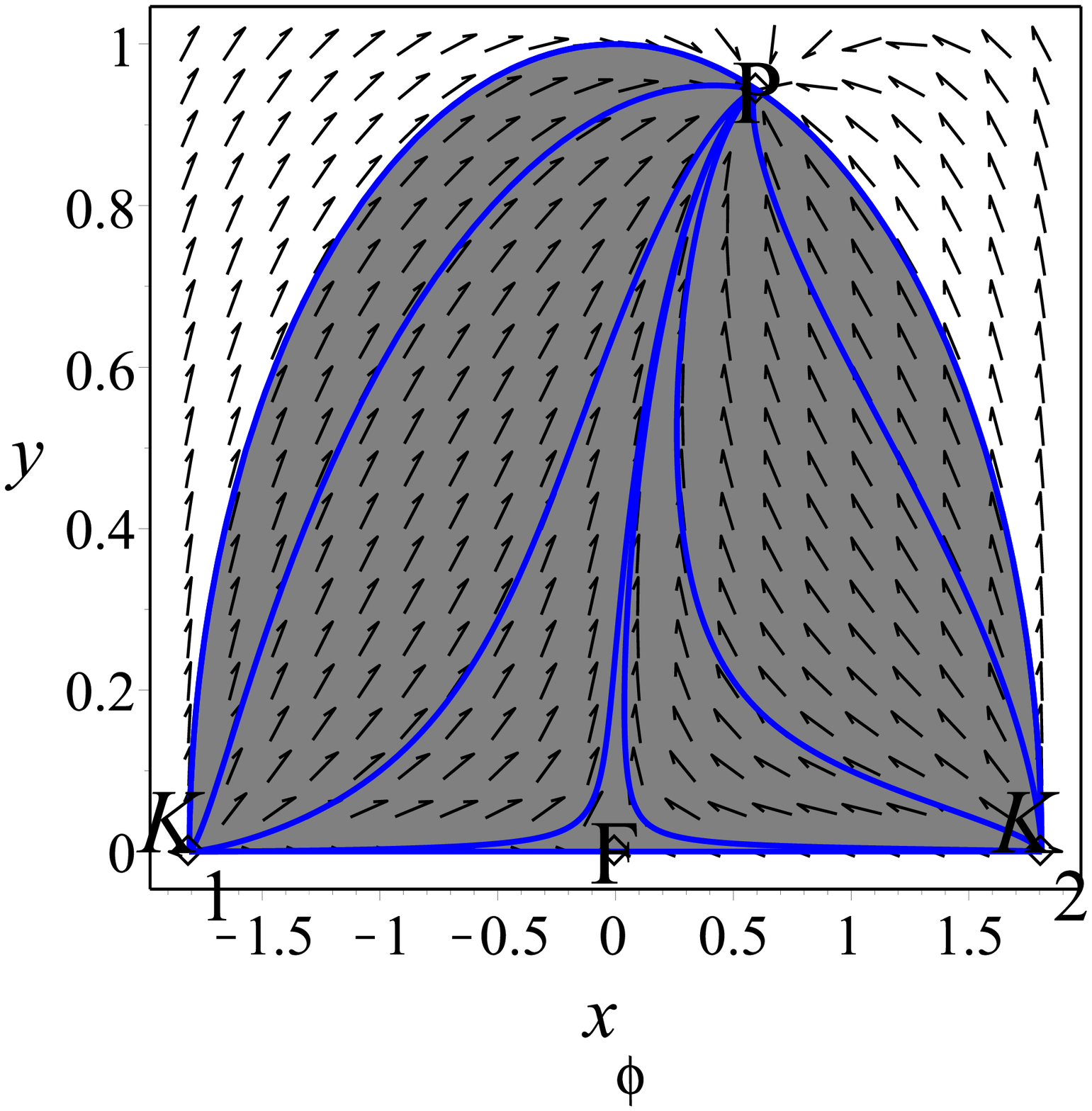}}
\subfigure[\;\;$m=0.9, n=0.2,\gamma=1$.]{\includegraphics[width=0.3\textwidth]{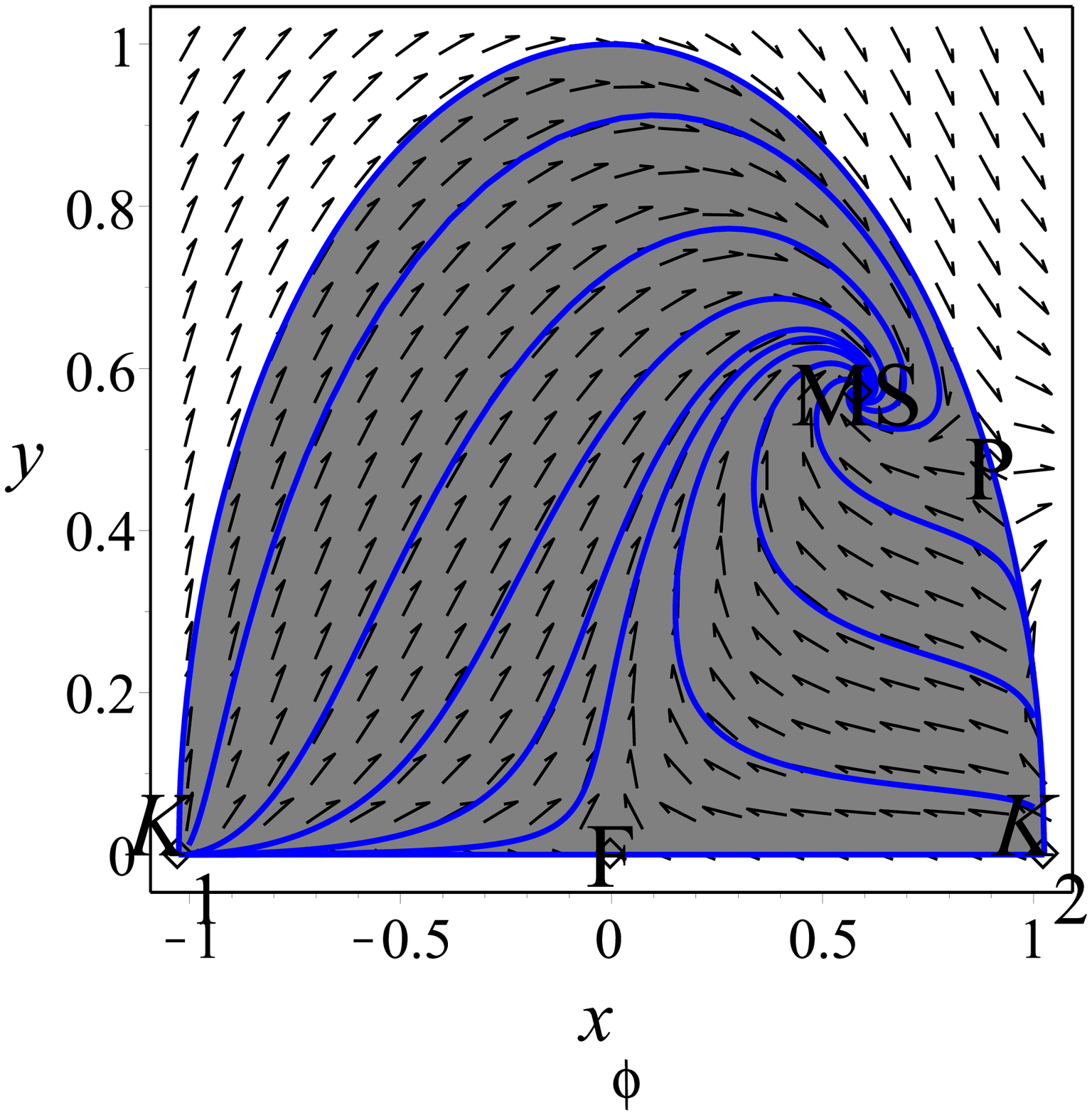}} 
\subfigure[\;\;$m=2, n=0.5,\gamma=1$.]{\includegraphics[width=0.3\textwidth]{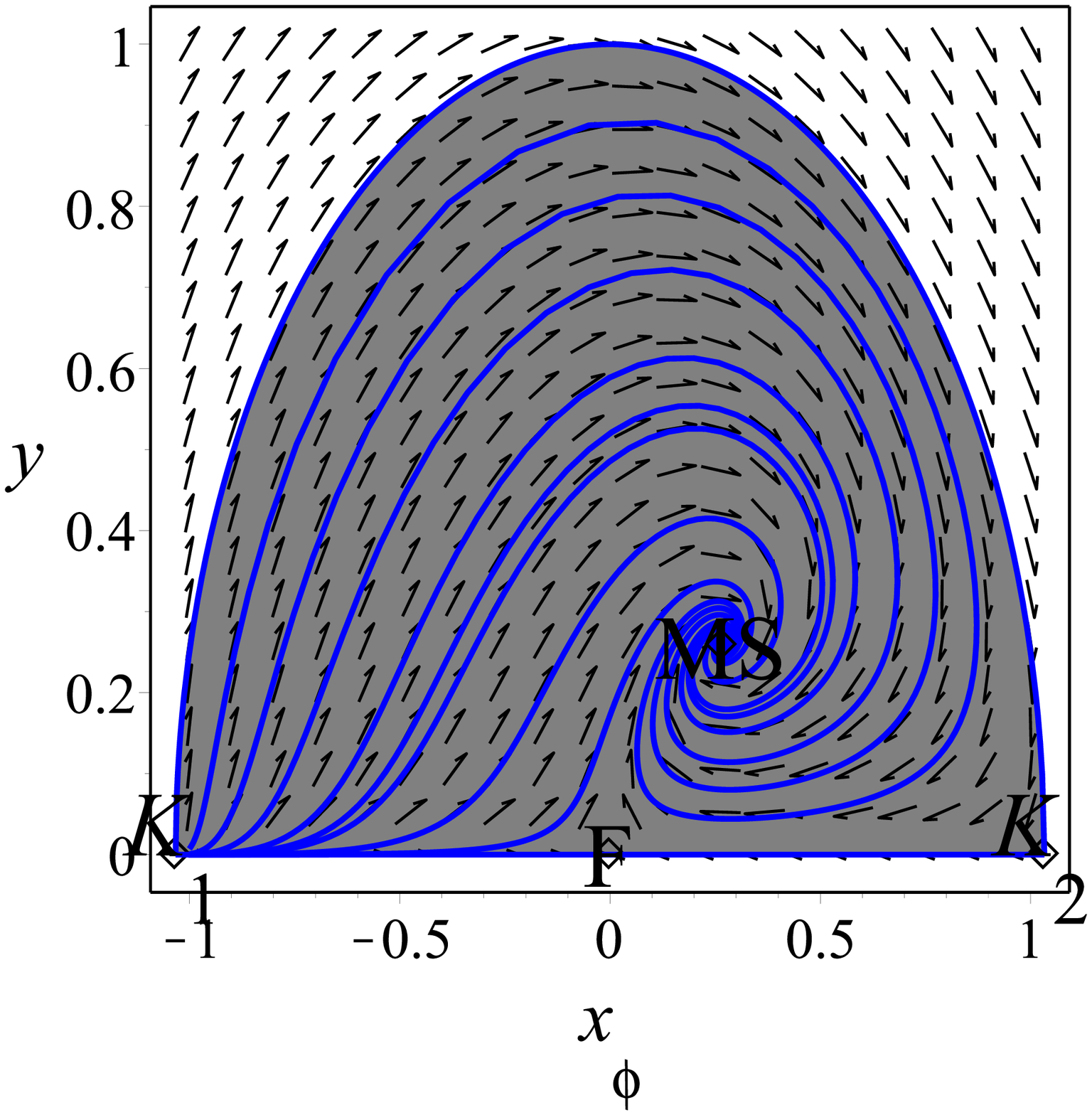}}
\subfigure[\;\;$m=0.9, n=0.2,\gamma=0.1$.]{\includegraphics[width=0.3\textwidth]{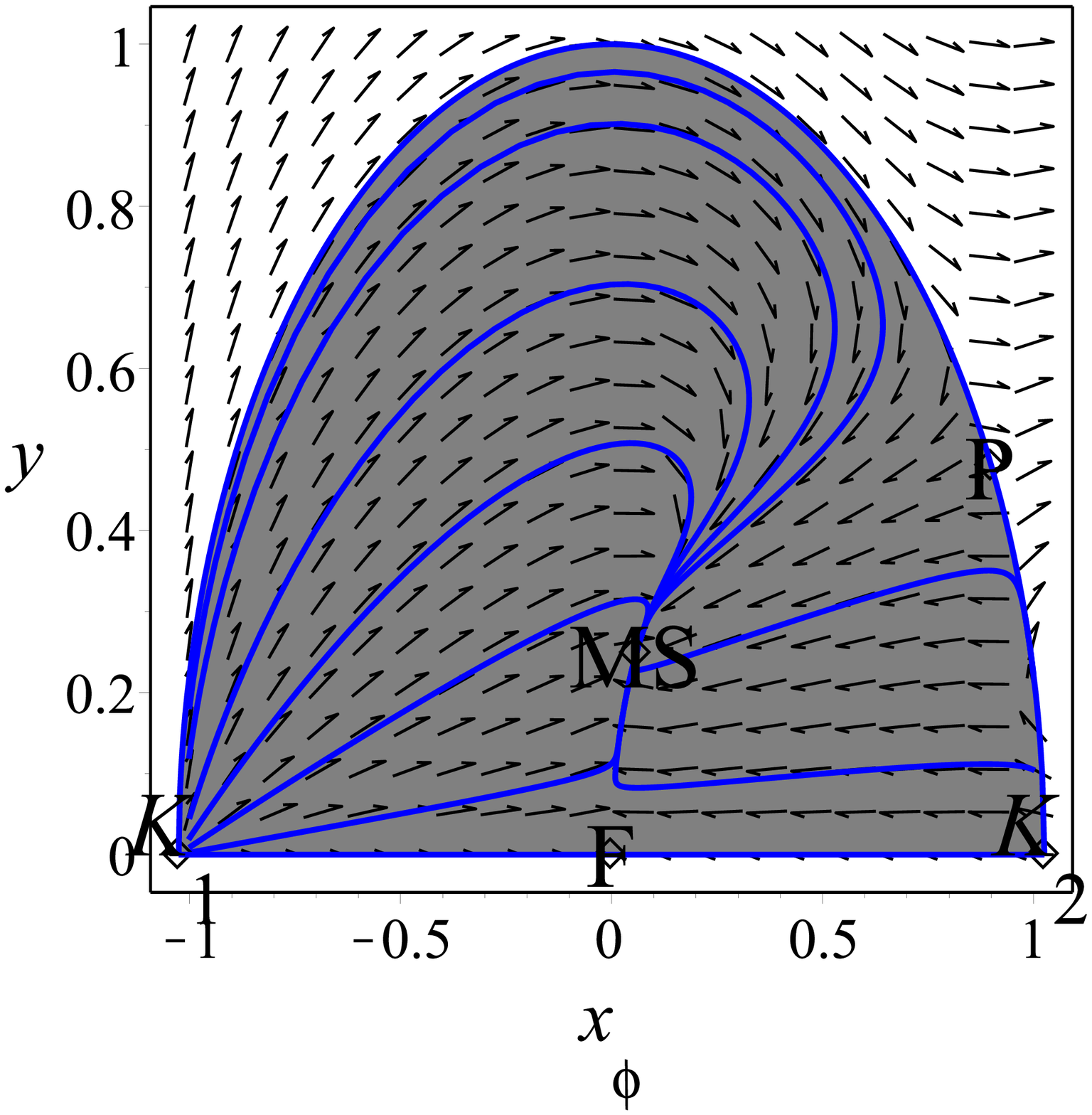}} 
\subfigure[\;\;$m=0.5, n=0.6,\gamma=1$.]{\includegraphics[width=0.3\textwidth]{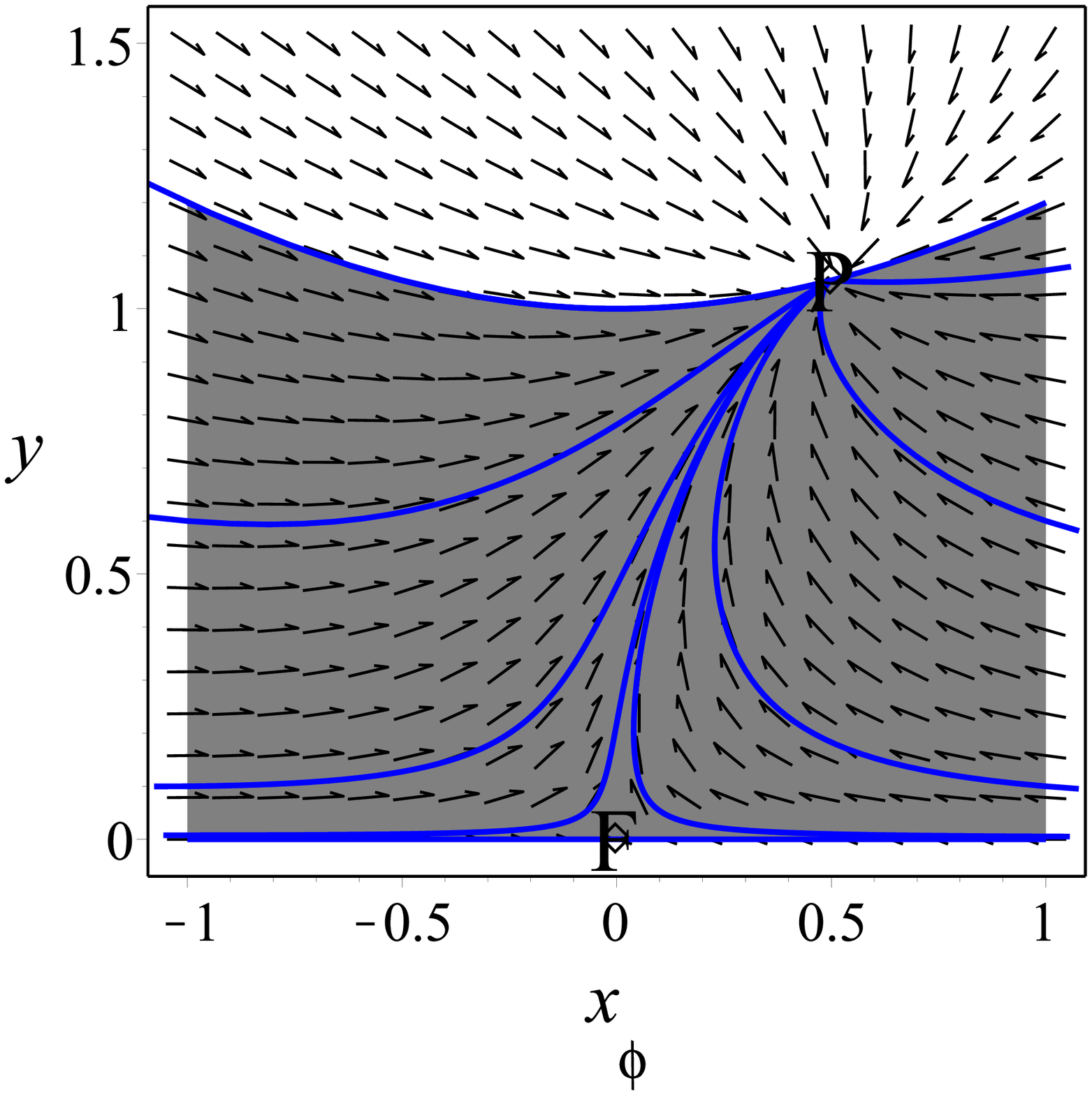}}
\caption{\label{FIG1} Some orbits in the phase plane of the system \eqref{eq13}-\eqref{eq14} for five choices of the parameters. }
\end{figure*}

In figure \ref{FIG1} it is shown some orbits of the system \eqref{eq13}-\eqref{eq14} for (a) $m=0.6, n=0.5, \gamma=1$,  (b)  $m=0.9, n=0.2, \gamma=1$, (c) $m=2.0, n=0.5, \gamma=1$, (d) $m=0.9, n=0.2, \gamma=0.1$ and (e) $m=0.5, n=0.6, \gamma=1$. The shadowed region represents the physical portion of the phase plane which is either half ellipse or half hyperbola.

\begin{table*}[t!]
\resizebox{\textwidth}{!}{
\begin{tabular}{|c|c|c|c|c|c|}
\hline
Label & $(x_\phi,x_\psi,y,z)$& Existence& Deceleration $q$& Curvature& Solution\\
\hline
$M$& $(0,0,0,0)$& All $m$ and $n$ & $q=0$& $\R<0$ & Milne\\[0.2cm]
$F$& $(0,0,0,1)$& All $m$ and $n$ & $q=-1+\frac{3\gamma}{2}$& $\R=0$ & Flat FRW\\[0.2cm]
$C_{\pm}$ & $\left(\pm\sqrt{1+{x_{\psi}^*}^2},x_{\psi}^*,0,0\right)$& All $m$ and $n$ & $q=2$&
$\R=0$ & Massless Scalar Field \\[0.2cm]
${P}$&
$(m,-n,\sqrt{1-\delta},0)$& $\delta< 1$ &$q=-1+3\delta$&$\R=0$& Powerlaw\\[0.2cm]
${MS}$& $\left(\ds\frac{\ds m\gamma}{\ds2\delta},-\ds\frac{\ds n\gamma}{\ds 2\delta},\ds\frac{\sqrt{(2-\gamma)\gamma}}{2\sqrt{\delta}},\ds\sqrt{1-\ds\frac{\ds\gamma}{\ds2\delta}}\right)$&
$\delta> \ds\frac{\ds\gamma}{\ds2}$&
$q=-1+\frac{3\gamma}{2}$&$\R=0$& Matter Scaling\\[0.2cm]
${CS}$& $\left(\ds\frac{\ds m}{\ds 3\delta},-\ds\frac{\ds n}{\ds3\delta},\ds\frac{\sqrt{2}}{3\sqrt{\delta}},0\right)$& $\delta>0$&
$q=0$&$\R\;\left\{ \begin{array}{cc} 
<0& \text{if}\; \delta>\sfrac{1}{3}\\[0.2cm]
=0& \text{if}\; \delta=\sfrac{1}{3}\\[0.2cm]
>0& \text{if}\; 0<\delta<\sfrac{1}{3}\\[0.2cm]
\end{array}\right. $& Curvature Scaling\\[0.4cm]
\hline
\end{tabular}}
\caption{Location, existence and some properties of the equilibrium points of the system \eqref{0eqxphi}, \eqref{0eqxvphi}, \eqref{0eqy} and \eqref{0eqz} for $z\geq 0$, and $y\geq 0.$ We use the notation $\delta=m^2-n^2.$\label{0tab1}}
\end{table*}

\begin{table*}[t!]
\begin{center}
\resizebox{\textwidth}{!}{
\begin{tabular}{|c|c|c|}
\hline
Name & Eigenvalues& Dynamical character\\
\hline
$M$& $\ds\left(-\sfrac{2}{3},-\sfrac{2}{3},\sfrac{1}{3},\sfrac{1}{6}(2-3\gamma)\right)$& no hyperbolic if $\gamma=\sfrac{2}{3};$ unstable otherwise \\[0.2cm]
$F$ & $\ds\left(-1+\sfrac{\gamma}{2},-1+\sfrac{\gamma}{2},\sfrac{\gamma}{2},-\sfrac{1}{3}(2-3\gamma)\right)$ & no hyperbolic if $\gamma=\sfrac{2}{3}$ or $\gamma=2;$ unstable otherwise \\[0.3cm]
$C_{\pm}$ & $\ds\left(0, \sfrac{4}{3},1-\sfrac{\gamma}{2},1-nx_{\psi}^*\mp
m\sqrt{1+{x_{\psi}^*}^2}\right)$ & no hyperbolic\\[0.3cm]
${P}$&$\left(-\sfrac{2}{3}+2\delta,-1+\delta,-1+\delta,
-\sfrac{\gamma}{2}+\delta\right)$& no hyperbolic if $\delta=\sfrac{1}{3}$ or $\delta=\sfrac{\gamma}{2};$ \\ [0.2cm] &&
stable if $\ds 0<\gamma\leq \sfrac{2}{3}$ and $\ds \delta<\sfrac{\gamma}{2}$, or $\ds \sfrac{2}{3}<\gamma<2$ and $\ds \delta<\sfrac{1}{3};$ \\ [0.2cm] && unstable otherwise
 \\[0.3cm]
${MS}$&
$\left(\ds -1+\frac{\gamma}{2},-\ds \frac{2}{3}+\gamma,\ds\frac{1}{4}\left(-2+\gamma\pm \sqrt{(2-\gamma)\left(2-9\gamma+\frac{4\gamma^2}{\delta}\right)}\right)\right)$& non hyperbolic if $\gamma=0$ or $\gamma=\sfrac{2}{3}$ or $\gamma=2;$ \\ [0.2cm] &&
unstable (saddle) if $\frac{2}{3}<\gamma<2;$ 
stable otherwise. \\[0.3cm]
${CS}$&
$\left(\ds -\frac{2}{3},\ds \frac{1}{6}(2-3\gamma),-\ds\frac{1}{3}\pm \sqrt{\frac{4}{3\delta}-3}\right)$& non hyperbolic if $\gamma=\sfrac{2}{3}$ or $\delta=\sfrac{3}{7};$ \\ [0.2cm] &&
stable when  $\frac{2}{3}<\gamma<2,\;\delta>\frac{3}{7};$ 
unstable otherwise.
\\[0.4cm]
\hline
\end{tabular}}
\end{center}
\caption{Eigenvalues, and dynamical character of the fixed points  of the system \eqref{0eqxphi}, \eqref{0eqxvphi}, \eqref{0eqy} and \eqref{0eqz} for $y\geq 0$ and $z\geq 0.$\label{0tab2}}
\end{table*}

\begin{table*}[t!]
\begin{center}
\resizebox{\textwidth}{!}{
\begin{tabular}{|l|c|c|c|c|c|}
\hline
	& $(y,x_\phi)$ 	& \multicolumn{2}{c|}{{\scriptsize Existence}} 		& {\scriptsize Eigenvalues} & {\scriptsize Dynamical character} \\ \hline
	&  	& {\scriptsize Case i)} $m^2\geq n^2$ & {\scriptsize Case ii)} $m^2< n^2$& \multicolumn{2}{c|}{}\\ \hline 
    $K_{1,2}$   & $\left(0,\mp\frac{m}{\sqrt{\delta}}\right)$ & $\delta>0$	&  $\nexists$ & $\left(1\mp\sqrt{\delta},2-\gamma\right)$ & $K_1:$ non-hyperbolic if  $\delta=1$\\
	& & & & & {saddle if $\delta>1$} \\
	& & & & & {source otherwise} \\
    & & & & &$K_2:$  source. \\[0.2cm]
\hline
$F$ 	& $(0,0)$	& \multicolumn{2}{c|}{$\forall m \forall n$}	& $\left(\sfrac{\gamma}{2}-1,\sfrac{\gamma}{2}\right)$ & saddle \\[0.2cm]\hline
$P$ & $\left(\sqrt{1-\delta},m\right)$ 	&  $\delta\leq 1$ &  $\forall m \forall n$ & $\left(\delta-1, 2\delta-\gamma\right)$ & non-hyperbolic if  $\delta=1$ or $\delta=\sfrac{\gamma}{2}$ \\
	& & & & & {a sink if $\delta <\sfrac{\gamma }{2}$}\\	
	& & & & & {saddle otherwise}  \\[0.2cm]	
\hline
$MS$ & $\left(\frac{1}{2}\sqrt{\sfrac{(2-\gamma)\gamma}{\delta}},\sfrac{m\gamma}{2\delta}\right)$ & $\delta\geq\sfrac{\gamma}{2}>0$	&  $\nexists$ & $\sfrac14\left(-2+\gamma\pm\sqrt{(2-\gamma)(2-9\gamma+\sfrac{4\gamma^2}{\delta})}\right)$ & non-hyperbolic if $\delta=\sfrac{\gamma}{2}$\\
	& & & & & {stable node if $0<\gamma\leq\sfrac29$ or } \\[0.1cm]
	& & & & & if $\sfrac29<\gamma\leq\sfrac23,\delta\leq \sfrac{4\gamma^2}{(9\gamma-2)}$  \\
	& & & & & stable spiral if \\
	& & & & & $\sfrac29<\gamma<2,\delta>\sfrac{4\gamma^2}{(9\gamma-2)}$\\
\hline
\end{tabular}}
\caption[Equilibrium points for $\delta=0$]{{\em The equilibrium
points of the system \eqref{sym_ds:1}-\eqref{sym_ds:2}. We use the notation $\delta=m^2-n^2.$ It is assumed $0<\gamma<2$. \label{i_table1}}}
\end{center}
\end{table*}

\begin{figure*}[t!]
\centering
\subfigure[\;\;$m=0.5, n=0.6, \gamma=1$.]{\includegraphics[width=0.3\textwidth]{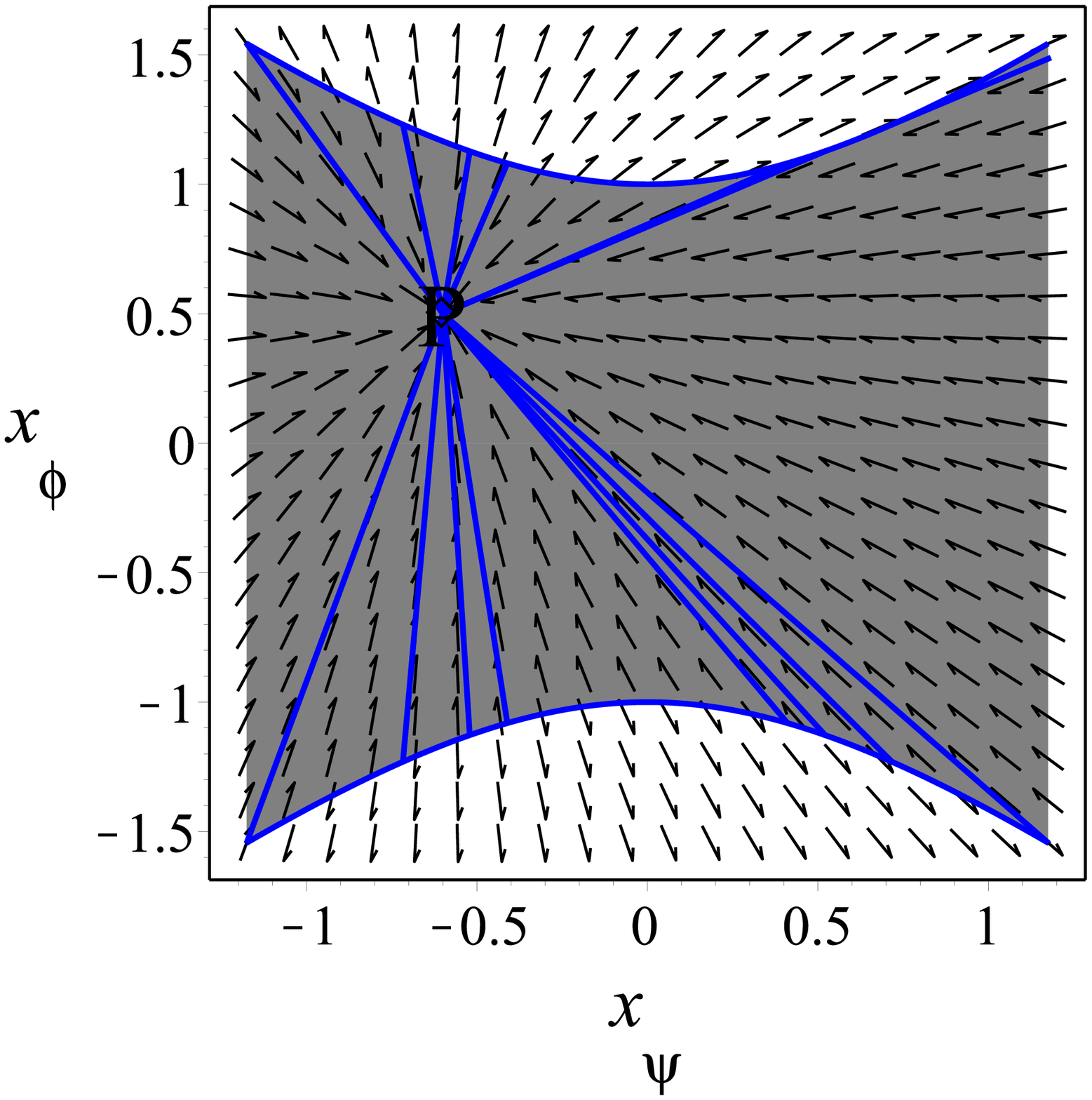}}
\subfigure[\;\;$m=0.6, n=0.5, \gamma=1$.]{\includegraphics[width=0.3\textwidth]{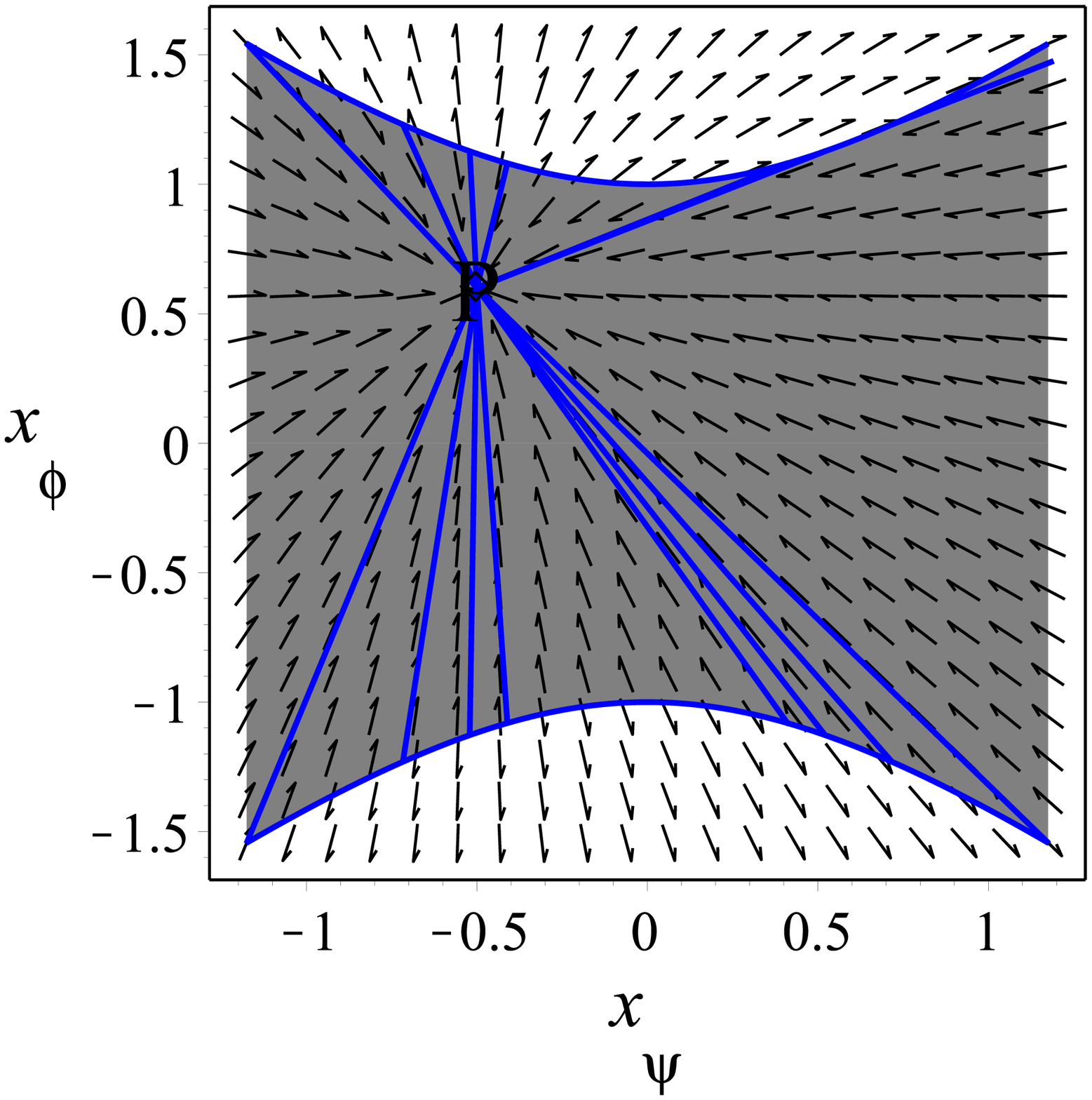}}
\centering
\subfigure[\;\;$m=0.2, n=0.2, \gamma=1$.]{\includegraphics[width=0.3\textwidth]{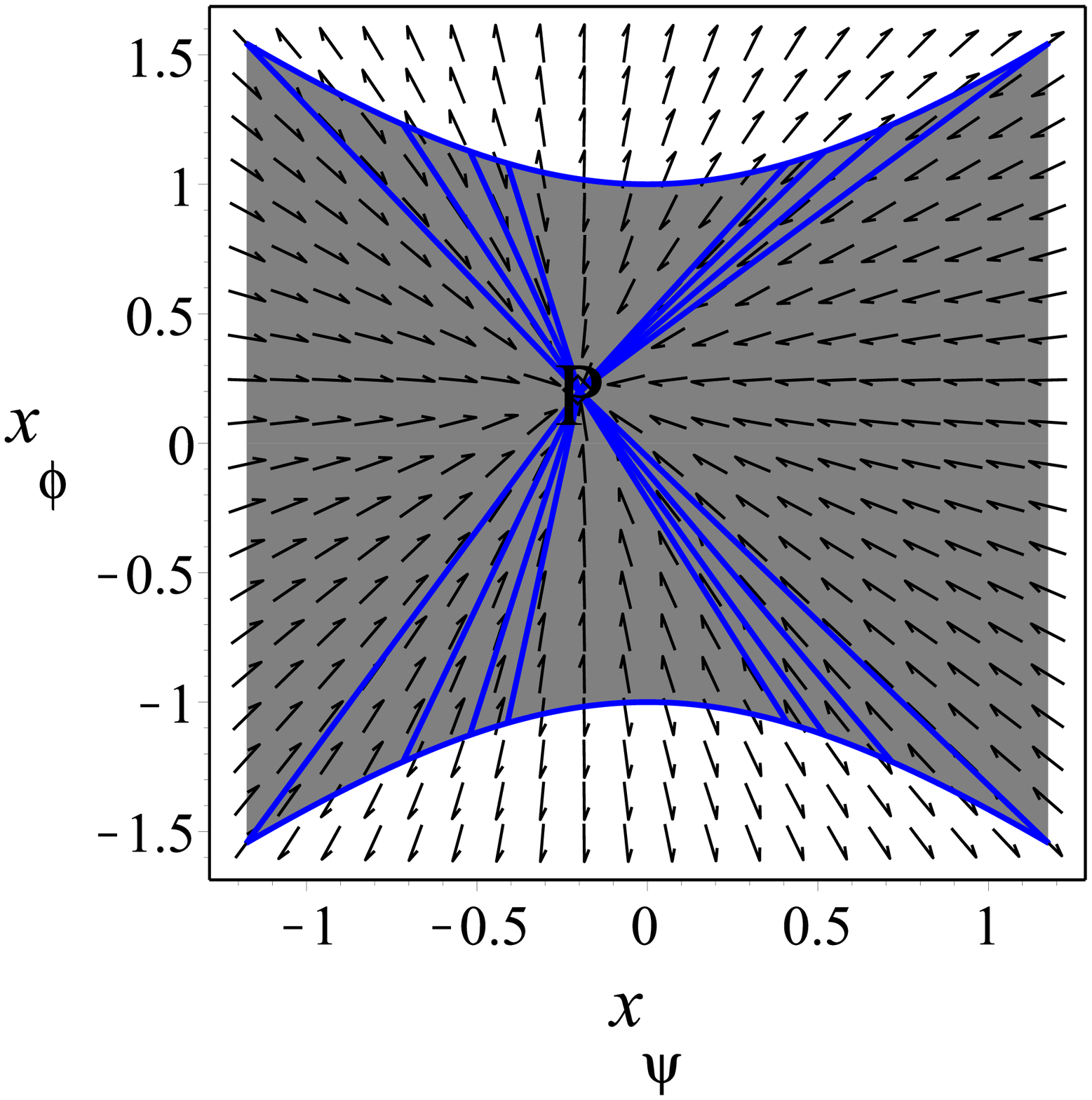}}
\caption{\label{FIG2} Some orbits in the phase plane of the system \eqref{sym_ds:1}-\eqref{sym_ds:2} for three choices of the parameters.}
\end{figure*}

\subsection{Flat FRW geometry and dust matter}\label{sectionIII}

For the flat FRW geometry and dust matter  dynamics is
governed by the vector field \cite{31}:
\begin{align}
&x_\phi'=\frac{1}{3} \left(3 m y^2+(q-2) x_\phi\right)\label{eqxphi}\\
&x_\vphi'=-\frac{1}{3} \left(3 n y^2-(q-2) x_\psi \right)\label{eqxvphi}\\
&y'=\frac{1}{3} (1+q-3(m x_\phi+n x_\psi)) y\label{eqy},
\end{align}
where the deceleration factor
is \be q=\frac{1}{2} \left(3
\left(x_\phi^2-x_\psi^2-y^2\right)+1\right),\ee
defined in the phase space given by \be\Psi=\{{\bf
x}=(x_\phi,x_\vphi,y):0\le
x_\phi^2-x_\vphi^2+y^2\le1\}.\label{phase_space}\ee

\begin{table}[ht]
\begin{center}
\begin{tabular}{|c|c|c|c|c|c|}\hline
Name &$x_\phi$&$x_\vphi$&$y$&Existence&$q$\\
\hline 
$O$& $0$& $0$& $0$& All $m$ and $n$ &$\sfrac{1}{2}$\\[0.2cm]
$C_{\pm}$ & $\pm\sqrt{1+{x_{\psi}^*}^2}$& $x_{\psi}^*$& $0$&
All $m$ and $n$  &$2$ \\[0.2cm]
${P}$&
$m$& $-n$ &$\sqrt{1-\delta}$&$\delta< 1$&$-1+3\delta$\\[0.2cm]
${T}$& $\ds\frac{\ds m}{\ds{2}{\delta}}$& $-\ds\frac{\ds
n}{\ds{2}{\delta}}$&
$\ds\frac{1}{2\sqrt{\delta}}$&$\delta\ge1/2$&$\ds\frac{1}{2}$\\[0.4cm]
\hline
\end{tabular}
\end{center}
\caption{Location, existence and deceleration factor of the
equilibrium points for $m>0$,  $n>0$ and $y>0$  \cite{31}.\label{table1}}
\end{table}

In table 
\ref{table1} it is presented some information of the equilibrium
points $O, C_{\pm}, T, P$ from \cite{31}. We have the following cases \cite{31}: 
Case i) If $m<\sqrt{n^2+1/2},$ the point $P$ is an stable node and $T$ does not exists.  Case ii) For  $\sqrt{n^2+1/2}<m\leq \sqrt{n^2+4/7},$ the point $T$ is a stable node and $P$ is
a saddle point. Case iii) For $\sqrt{n^2+4/7}<m<\sqrt{1+n^2},$ point T is a spiral node and the point P is a saddle.   Case iv) For  $m>\sqrt{1+n^2}$ the point $T$ is a spiral node whereas the point $P$ does not exist. 

For $\R=0, z=0$, that is, the quintom cosmological model (%
\ref{expquintomlag}) with potential (\ref{ExponentialPot}) in the spatially
flat geometry and without any other matter source, the dynamical system is reduced to 
\begin{align}
&x_\phi'=(x_{\phi}-m) \left(x_{\phi}^2-x_\psi^2-1\right),\label{sym_ds:1}\\
&x_\psi'=(n+x_\psi) \left(x_{\phi}^2-x_\psi^2-1\right),\label{sym_ds:2}
\end{align}
defined on the phase plane
\begin{equation}
\{(x_\phi, x_\psi)\in \mathbb{R}^2: x_{\phi}^2-x_\psi^2\leq 1\}.
\end{equation}
In the figure \ref{FIG2} are presented some orbits in the phase plane of the system \eqref{sym_ds:1}-\eqref{sym_ds:2} for three choices of the parameters. This figures depict the typical behavior of the quintom model with exponential potential and no other matter component (vacuum quintom), where the $P$ is always an stable node. The shadowed region corresponds to the physical region. The region $w<-1$ is reached provided $m^2-n^2<0$, so the crossing does indeed occur in the case $m=0.5, n=0.6$ whence $P$ is a phantom-like solution. For $m=0.6, n=0.5$ the equation of state remains above $w=-1$ and $P$ is a quintessence-like solution, and for $m=n=0.2$, $w=-1$, such that $P$ represents a de Sitter solution.

\begin{rem}
For $\R=0, z=0$, the sets
 $\{{\bf x}\in\mathbb{R}^2: m x_\phi+n x_\psi=0\}$ and
$\{{\bf x}\in\mathbb{R}^2:\frac{x_\phi-x_\psi}{m+n}-1=0\}$ are invariant sets of the flow of \eqref{sym_ds:1}-\eqref{sym_ds:2}.
\end{rem}

Defining $\Psi_1=m x_\phi+n x_\psi$, we show that 
$\Psi_1'=\Psi_1 \left(x_{\phi}^2-x_\psi^2-1\right)$. Such that $\{{\bf x}: \Psi_1=0\}$ is an invariant set. 
Defining  $\Psi_2=\frac{x_\phi-x_\psi}{m+n}-1$, we show that 
$\Psi_2'=\Psi_2 \left(x_{\phi}^2-x_\psi^2-1\right)$. Such that $\{{\bf x}: \Psi_2=0\}$ is an invariant set.

\subsubsection{Normal expansion up to arbitrary order}\label{normal}

In this section we
obtain the normal form expansion for an arbitrary point lying on the
curve $C_-$  \cite{Leon:2008aq}. 
\begin{prop}\label{prop4}
Let be the vector field ${\bf X}$ given by
(\ref{eqxphi}-\ref{eqy}) which is $C^\infty$ in a neighborhood of
${\bf x}^*=(x_{\phi}^*,x_{\psi}^*,y^*)^T\in C_-.$ Let $m\geq
n>0,$ and $x_{\psi}^*\in\mathbb{R},$ such that $\lambda_3=1-n
x_\psi^*+m\sqrt{1+{x_\psi^*}^2}\notin \mathbb{Z}$, then, there exist
constants $a_r,\, b_r,\, c_r,\; r\geq 2$ (non necessarily
different from zero) and a transformation of coordinates ${\bf
x}\rightarrow {\bf y},$ such that (\ref{eqxphi}-\ref{eqy}) has
normal form
\ben
y_1'&=&\sum_{r=2}^N a_r y_1^r+{O}(|{\bf y}|^{N+1}),\label{N1}\\
y_2'&=&y_2\left(1+\sum_{r=2}^N b_r y_1^{r-1}\right)+{O}(|{\bf y}|^{N+1}),\label{N2}\\
y_3'&=&y_3\left(\lambda_3+\sum_{r=2}^N c_r
y_1^{r-1}\right)+{O}(|{\bf y}|^{N+1}),\label{N3} \een defined in
neighborhood of ${\bf y}=(0,0,0).$
\end{prop}
{\bf Proof}. Applying a linear coordinate transformation the linear part of ${\bf X}$ is reduced to $${\bf
X}_1({\bf x})= \left(\begin{array}{ccc}
0 & 0 & 0 \\
0 & 1 & 0 \\
0 & 0 & \lambda_3
\end{array}\right)\left(\begin{array}{c}
x_1\\
x_2 \\
x_3
\end{array}\right)= {\bf J}{\bf x}.$$ By the hypothesis $m\geq n>0 \implies\lambda_3^->1\; \forall x_{\psi}^*\in\mathbb{R}.$ Therefore, the eigenvalues of ${\bf J}$
are different and ${\bf J}$ is diagonal; then, the corresponding
eigenvectors
$B=\left\{{\bf x}^{\bf m}{\bf e}_i:=x_1^{m_1} x_2^{m_2} x_3^{m_3}{\bf e}_i | m_j\in\mathbb{N}, \sum m_j=r, i,j=1,2,3\right\}$
form a basis of $H^r.$ Let denote $\lambda_1=0, \lambda_2=1, \lambda_3=1-n
x_\psi^*+m\sqrt{1+{x_\psi^*}^2}$. 
Let ${\bf
L_J}$ be the linear operator that assigns to ${\bf h(y)}\in H^r$
the Lie bracket of the vector fields ${\bf J y}$ and ${\bf h(y)}$:
\begin{align} {\bf L_J}: H^r& \rightarrow H^r\nonumber\\
     {\bf h}   & \rightarrow  {\bf L_J} {\bf h (y)}={\bf D h(y)} {\bf J y}- {\bf J h(y)}.
\end{align}

Applying this operator to monomials ${\bf x}^{\bf m}{\bf e}_i$, where $m$ is a multiindex of order r and ${\bf e}_i$ basis vector of $\mathbb{R}^3$, we find
$${\bf L}_{\bf J} {\bf x}^{\bf m}{\bf e}_i
=\left\{({\bf m}\cdot {\bf \lambda})-\lambda_i\right\}
{\bf x}^{\bf m}{\bf e}_i.$$ The eigenvectors in $B$ for which
$\Lambda_{{\bf m},i}\equiv ({\bf m}\cdot {\bf \lambda})-\lambda_i\neq 0$
form a basis of $B^r={\bf L}_{\bf J}(H^r)$ and those such that $\Lambda_{{\bf m},i}=0,$  associated to the resonant
eigenvalues, form a basis for the
complementary subspace, $G^r,$ of $B^r$ in $H^r.$
\\
Since $\lambda_1=0,$ and $\lambda_3\notin \mathbb{Z}$ by hypothesis, the resonant equations of order $r$ has a unique solution:
\begin{align}
&m_2+\lambda_3 m_3=0 \Rightarrow m_1=r, m_2=m_3=0,\\
&m_2+\lambda_3 m_3=1 \Rightarrow m_1=r-1, m_2=1, m_3=0,\\
& m_2+\lambda_3 m_3=\lambda_3 \Rightarrow m_1=r-1, m_2=0, m_3=1
\label{hom3}. \end{align}

Then, $\left\{x_1^r {\bf e}_1, x_1^{r-1} x_2 {\bf e}_2,x_1^{r-1}
x_3 {\bf e}_3\right\},$ form a basis for  $G^r$ in $H^r.$

By applying theorem \ref{NFTheorem}, we have that there exists a
coordinate transformation ${\bf x}\rightarrow {\bf y},$ such that
(\ref{eqxphi}-\ref{eqy}) has normal form (\ref{N1}-\ref{N3}) where
$a_r, b_r$ and $c_r$ are some real constants. The values of of all these constants can be uniquely determined,
up to the desired order, by inductive construction in $r.$ $\blacksquare$
\\Now we integrate the truncated system \eqref{N1}, \eqref{N2}  \eqref{N3} up to order $N$
with initial condition $\left(y_1(t_0),y_2(t_0),y_3(t_0)\right)=(y_{10},y_{20},y_{30}).$
\begin{enumerate}
\item If $y_{10}=0,$ then $y_1(t)=0,$  $y_2(t)=y_{20}e^{\tau-\tau_0}$ and
$y_3(t)=y_{30}e^{\lambda_3\left(\tau-\tau_0\right)}$ for all
$t\in \mathbb{R}.$ Then, the orbit approach the origin as $\tau\rightarrow-\infty$ provided $\lambda_3>0.$
\item 
If $y_{10}\neq 0,$ we obtain the quadratures 
\begin{align}
&\tau-\tau_0=\int_{y_{10}}^{y_1} \left(\sum_{r=2}^N a_r \zeta^r\right)^{-1} \mathrm{d}\zeta,\label{tauy1}\\
&y_2(\tau)=y_{20} e^{\tau-\tau_0} \prod_{r=2}^N \exp\left[b_r\int_{\tau_0}^\tau y_1(t)^{r-1} \mathrm{d} t\right],\label{eqNy2}\\
&y_3(\tau)=y_{30}e^{\lambda_3\left(\tau-\tau_0\right)} \prod_{r=2}^N
\exp\left[c_r\int_{\tau_0}^\tau y_1(t)^{r-1} \mathrm{d}
t\right].\label{eqNy3} \end{align}
\end{enumerate}
The $y_1$-component of the orbit passing through
$(y_{10},y_{20},y_{30})$ at $\tau=\tau_0$ with $y_{10}\neq 0$ is
obtained by inverting the quadrature (\ref{tauy1}).

The other components are given by

\ben
y_2=y_{20} \exp\left[\int_{y_{10}}^{y_1}\frac{1+\sum_{r=2}^N b_r \zeta^{r-1}}{\sum_{r=2}^N a_r \zeta^r}\mathrm{d} \zeta\right],\\
y_3=y_{30} \exp\left[\int_{y_{10}}^{y_1}\frac{1+\sum_{r=2}^N c_r
\zeta^{r-1}}{\sum_{r=2}^N a_r \zeta^r} \mathrm{d} \zeta\right].
\een

\subsubsection{Normal expansion to third order for
$C_-$}

In this section we show normal form expansions for the vector
field (\ref{eqxphi}-\ref{eqy}) defined in a vicinity of $C_-$
expressed in the form of proposition \ref{Prop2.5} (see \cite{Leon:2008aq})
\begin{prop}\label{Prop2.5} Let be the vector field ${\bf X}$ given by
(\ref{eqxphi}-\ref{eqy}) which is $C^\infty$ in a neighborhood of
${\bf x}^*=(x_{\phi}^*,x_{\psi}^*,y^*)^T\in C_-.$ Let $m\geq
n>0,$ $x_{\psi}^*\in\mathbb{R},$ such that $\lambda_3^-=1-n
x_\psi^*+m\sqrt{1+{x_\psi^*}^2}\notin\mathbb{Z}$, then, there exist
a transformation to new coordinates $x\rightarrow z,$ such that
(\ref{eqxphi}-\ref{eqy}), defined in a vicinity of ${\bf x}^*,$
has normal form

\ben \label{2_17} z_1'&=&O(|z |^4),
\\
\label{2_18} z_2' &=& z_2+O(|z |^4),
\\
\label{2_19} z_3'&=&\left(\lambda^-_3+c_2 z_1+c_3
z_1^2\right)z_3+O(|z |^4),\een where  $c_2=-n+\frac{m
x_\psi^*}{\sqrt{1+{x_\psi^*}^2}}$ and $c_3=-\frac{n
x_\psi^*}{2\left(1+{x_\psi^*}^2\right)}+\frac{m}{2\sqrt{1+{x_\psi^*}^2}}.$

\end{prop}
{\bf Proof}. \\
\bigskip
\textbf{First step: transforming to the Jordan form}

Using a linear coordinate transformation the system
(\ref{eqxphi}-\ref{eqy}) is transformed to 
\be {\bf x}'={\bf J} {\bf
x}+{\bf X}_2({\bf x})+{\bf X}_3({\bf x})\label{Jordan3}\ee
where ${\bf x}$ stands for the phase vector ${\bf x}=\left(x_1,\,x_2,\,x_3\right)^T,$ 
\be {\bf J}=\left(\begin{array}{ccc} 0 & 0 & 0\\
                0 & 1 & 0\\
                0 & 0 & 1-n {x_{\psi}^*}+m \sqrt{1+{x_{\psi}^*}^2}\end{array}\right),\label{JordanMatrix}\ee
                \\
{\small \be {\bf X}_2({\bf x})=\left(\begin{array}{c} X_{(1,1,0),1}{x_1
   \,x_2}+X_{(0,0,2),1} x_3^2\vspace{10pt}\\
               X_{(2,0,0),2} x_1^2+ X_{(0,2,0),2} x_2^2+ X_{(0,0,2),2} x_3^2 \vspace{10pt}\\
                X_{(1,0,1),3} x_1  x_3+   X_{(0,1,1),3} x_2  x_3\end{array}\right),\label{X2}\ee
                }\\with\\ ${X}_{(1,1,0),1}=\frac{1}{{x_{\psi}^*}}$, ${X}_{(0,0,2),1}=m  {x_{\psi}^*}\sqrt{1+{x_{\psi}^*}^2}-n
\left(1+{x_{\psi}^*}^2\right)$, ${X}_{(2,0,0),2}=-\frac{{x_{\psi}^*}}{2 \left({x_{\psi}^*}^2+1\right)}$, ${X}_{(0,2,0),2}=\frac{3}{2 {x_{\psi}^*}}$, ${X}_{(0,0,2),2}=\frac{1}{2} {x_{\psi}^*}
\left(-2 m
   \sqrt{1+{x_{\psi}^*}^2}+2 n {x_{\psi}^*}-1\right)$, 
${X}_{(1,0,1),3}=-n+\frac{m {x_{\psi}^*}}{\sqrt{1+{x_{\psi}^*}^2}},$ ${X}_{(0,1,1),3}=-\frac{
   \left(-m \sqrt{1+{x_{\psi}^*}^2}+n {x_{\psi}^*}-1\right)}{{x_{\psi}^*}}$; \\and
{\small
\be{\bf X}_3({\bf x})=\left(\begin{array}{c} X_{(3,0,0),1}{x_1^3}+X_{(1,2,0),1} x_1 x_2^2+X_{(1,0,2),1} x_1 x_3^2\vspace{10pt}\\
               X_{(2,1,0),2} x_1^2 x_2+ X_{(0,3,0),2} x_2^3 +X_{(0,1,2),2} x_2 x_3^2\vspace{10pt}\\
                X_{(2,0,1),3} x_1^2 x_3 +  X_{(0,2,1),3} x_2^2 x_3 +  X_{(0,0,3),3} x_3^3
                \end{array}\right),\label{X3}\ee 
                }\\
                with\\ ${X}_{(3,0,0),1}=-\frac{1}{2 \left({x_\psi^*}^2+1\right)}$, ${X}_{(1,2,0),1}=\frac{1}{2 {x_\psi^*}^2}$, ${X}_{(1,0,2),1}=-\frac{1}{2}$, 
${X}_{(2,1,0),2}=-\frac{1}{2 \left({x_\psi^*}^2+1\right)}$, ${X}_{(0,3,0),2}=\frac{1}{2 {x_\psi^*}^2}$, ${X}_{(0,1,2),2}=-\frac{1}{2}$, 
${X}_{(2,0,1),3}=-\frac{1}{2 \left({x_\psi^*}^2+1\right)}$, ${X}_{(0,2,1),3}=\frac{1}{2 {x_\psi^*}^2}$, ${X}_{(0,0,3),3}=-\frac{1}{2}$.
\newpage

\textbf{Second step: simplifying the quadratic part}

By the hypotheses the eigenvalues
$\lambda_1=0,\lambda_2=1,\lambda_3^-=1-n {x_{\psi}^*}+m
\sqrt{1+{x_{\psi}^*}^2}$ of ${\bf J}$ are different. Hence, its
eigenvectors form a basis of $\mathbb{R}^3.$ The linear operator
$${\bf L}^{(2)}_{\bf J}: H^2\rightarrow H^2$$ has eigenvectors ${\bf
x}^{\bf m}{\bf e}_i$ with eigenvalues $\Lambda_{{\bf
m},i}=m_1\lambda_1+m_2\lambda_2+\lambda_3 m_3-\lambda_i,$
$i=1,2,3,$ $m_1,m_2,m_3\geq 0,$ $m_1+m_2+m_3=2.$ The eigenvalues
$\Lambda_{{\bf m},i}$ for the allowed ${\bf m},i$ are 
$\Lambda_{(1,1,0),1}=1$, $\Lambda_{(0,0,2),1}=2\left(1-n {x_{\psi}^*}+m \sqrt{1+{x_{\psi}^*}^2}\right)$, $\Lambda_{(2,0,0),2}=-1$, $\Lambda_{(0,2,0),2}=1$, $\Lambda_{(0,0,2),2}=1-2 n {x_{\psi}^*}+2 m \sqrt{1+{x_{\psi}^*}^2}$, $\Lambda_{(1,0,1),3}=0$, $\Lambda_{(0,1,1),3}=1$. 

To obtain the normal form of
(\ref{2_17}-\ref{2_19}) wee must look for resonant terms, i.e.,
those terms of the form ${\bf x}^{\bf m}{\bf e}_i$ with ${\bf m}$
and $i$ such that $\Lambda_{{\bf m},i}=0$ for the available ${\bf
m},i.$ Only one term of second order is resonant :
$\Lambda_{(1,0,1),3}=0\rightarrow c_2 y_1 y_3 {\bf e}_3.$

The required function $${\bf h}_2: H^2\rightarrow H^2$$ to
eliminate the non-resonant quadratic terms is given by 
{\small \be {\bf
h}_2({\bf y})=\left(\begin{array}{c}
\frac{X_{(1,1,0),1}}{\Lambda_{(1,1,0),1}}{y_1
   \,y_2}+\frac{X_{(0,0,2),1}}{\Lambda_{(0,0,2),1}} y_3^2\vspace{10pt}\\
               \frac{X_{(2,0,0),2}}{\Lambda_{(2,0,0),2}} y_1^2+ \frac{X_{(0,2,0),2}}{\Lambda_{(0,2,0),2}} y_2^2+ \frac{X_{(0,0,2),2}}{\Lambda_{(0,0,2),2}} y_3^2 \vspace{10pt}\\
                 \frac{X_{(0,1,1),3}}{\Lambda_{(0,1,1),3}} y_2  y_3\end{array}\right),\label{h2}\ee}

The quadratic transformation \be {\bf x}\rightarrow {\bf y}+{\bf
h}_2(\bf y)\label{qtransform}\ee with ${\bf h}_2$ defined as in
(\ref{h2}) is the coordinate transformation required in theorem
\ref{NFTheorem}. By applying this theorem we prove the existence
of the required constant $c_2.$
Now, let us calculate the value of $c_2.$

By applying the transformation (\ref{qtransform}) the vector field
(\ref{Jordan3}) transforms to \be {\bf y}'={\bf J y}-{\bf
L}^{(2)}_{\bf J} {\bf h}_2 ({\bf y})+{\bf X}_2(\bf y)+\tilde{{\bf
X}}_3(\bf y)+{O}(|{\bf y}|^4),\label{Jordan4}\ee

Since \be -{\bf L}^{(2)}_{\bf J} {\bf h}_2 ({\bf y})+{\bf
X}_2({\bf y})= X_{(1,0,1),3}y_1 y_3 {\bf e}_3,\ee we have \be {\bf
y}'={\bf J y}+X_{(1,0,1),3}y_1 y_3 {\bf e}_3+\tilde{{\bf X}}_3(\bf
y)+{O}(|{\bf y}|^4),\ee i.e., $c_2=X_{(1,0,1),3}=-n+\frac{m
{x_{\psi}^*}}{\sqrt{1+{{x_{\psi}^*}}^2}}.$

The vector field $\tilde{{\bf X}}_3(\bf y)$ introduced above has
the coefficients:

{\footnotesize
\begin{eqnarray}
&{\tilde{X}}_{{ (2,0,1),3}}=\frac{m}{2
\sqrt{{x_\psi^*}^2+1}}-\frac{n {x_\psi^*}}{2  \left({x_\psi^*}^2+1\right)},\nonumber\\
&   {\tilde{X}}_{{ (1,2,0),1}}=\frac{3}{{x_\psi^*}^2}
   ,\nonumber\\
&   {\tilde{X}}_{{ (1,0,2),1}}=-\frac{n^2 \delta ^2+m\left(\delta +m\left(\delta ^2-1\right)\right)-n
   {x_\psi^*}[2 m \delta+1]+1}{\lambda
   _3^-}
   ,\nonumber\\
&   {\tilde{X}}_{{ (1,0,2),2}}=\frac{{x_\psi^*}\left[2 {x_\psi^*}
   m^2+\frac{\left({x_\psi^*}-2 \left(2 \left(\delta
   ^2-1\right) n+n\right)\right) m}{\delta }+n\left(2 n {x_\psi^*}-1\right)\right]}
   {2 {\lambda_3^-}},\nonumber\\
 &     {\tilde{X}}_{{ (0,3,0),2}}=\frac{5}{{x_\psi^*}^2},\nonumber\\
 &     {\tilde{X}}_{{ (0,2,1),3}}=\frac{\left(-\text{m$\delta $}+n {x_\psi^*}-2\right)
   \left(-2 \text{m$\delta $}+2 n {x_\psi^*}-3\right)}
   {2 {x_\psi^*}^2},\nonumber\\
&   {\tilde{X}}_{{
(0,1,2),1}}=\frac{\delta \left(4 \delta ^2 {x_\psi^*} m^3+4
   \delta  \Delta _1 m^2+\Delta _2 m+\text{n$\delta $} \Delta
   _3\right)}{2 {\lambda_3^-} {x_\psi^*}},\nonumber \\
&   {\tilde{X}}_{{ (0,1,2),2}}=-2 \left(\delta ^2-1\right) n^2+{x_\psi^*}[4
   \text{m$\delta $}+3] n-\text{m$\delta $}(2 \text{m$\delta $}+3)-3,\nonumber\\
&   {\tilde{X}}_{{
(0,0,3),3}}=-4 n {x_\psi^*}\left[n{x_\psi^*}-2\right]-5,
\end{eqnarray}
}

where $\delta =\sqrt{{x_\psi^*}^2+1},$ ${\Delta }_{{ 1}}=2
{x_\psi^*}-n\left(3 {x_\psi^*}^2+1\right),$ ${\Delta }_{{
2}}=4 n\left(-4 {x_\psi^*}^2+n\left(3
   {x_\psi^*}^2+2\right)
   {x_\psi^*}-2\right)+5 {x_\psi^*},$ ${\Delta }_{{ 3}}=-4 n {x_\psi^*}
   \left[n {x_\psi^*}-2\right]-5.$
\bigskip

\textbf{Third step: simplifying the cubic part}

After the last two steps, the equation (\ref{Jordan3}) is
transformed to \be {\bf y}'={\bf J y}+c_2 y_1 y_3 {\bf
e}_3+\tilde{{\bf X}}_3({\bf y})+{\bf O}(|{\bf y}|^4).\ee

As we will see later on, there is only one term of order three which is resonant: $\Lambda_{(2,0,1),3}=0\rightarrow c_3 z_1^2 z_3 {\bf
e}_3.$
Therefore, as in the last step, in order to eliminate non-resonant terms of
third order we will consider the coordinate transformation ${\bf
y}\rightarrow {\bf z}$ given by \be {\bf y}={\bf z}+{\bf h}_3
({\bf z})\label{transfz}\ee where
$${\bf h}_3: H^3\rightarrow
H^3$$  is defined by \be {\bf h}_3({\bf z})=\left(
\begin{array}{l}
\frac{{\tilde{X}}_{{\rm (1,2,0),1}}}{{\Lambda }_{{\rm (1,2,0),1}}}z_{{\rm 1}}z^{{\rm 2}}_{{\rm 2}}{\rm +}\frac{{\tilde{X}}_{{\rm (1,0,2),1}}}{{\Lambda }_{{\rm (1,0,2),1}}}z_{{\rm 1}}z^{{\rm 2}}_{{\rm 3}}{\rm +}\frac{{\tilde{X}}_{{\rm (0,1,2),1}}}{{\Lambda }_{{\rm (0,1,2),1}}}z_{{\rm 2}}z^{{\rm 2}}_{{\rm 3}} \\
\frac{{\tilde{X}}_{{\rm (1,0,2),2}}}{{\Lambda }_{{\rm (1,0,2),2}}}z_{{\rm 1}}z^{{\rm 2}}_{{\rm 3}}{\rm +}\frac{{\tilde{X}}_{{\rm (0,3,0),2}}}{{\Lambda }_{{\rm (0,3,0),2}}}z^{{\rm 3}}_{{\rm 2}}{\rm +}\frac{{\tilde{X}}_{{\rm (0,1,2),2}}}{{\Lambda }_{{\rm (0,1,2),2}}}z_{{\rm 2}}z^{{\rm 2}}_{{\rm 3}} \\
\frac{{\tilde{X}}_{{\rm (0,2,1),3}}}{{\Lambda }_{{\rm
(0,2,1),3}}}z^{{\rm 2}}_{{\rm 2}}z_{{\rm 3}}{\rm
+}\frac{{\tilde{X}}_{{\rm (0,0,3),3}}}{{\Lambda }_{{\rm
(0,0,3),3}}}z^{{\rm 3}}_{{\rm 3}} \end{array}
\right),\label{h3}\ee where 
$\Lambda_{(1,2,0),1}=2$, $\Lambda_{(1,0,2),1}=2 \left(m\sqrt{1+{x_\psi^*}^2}-n
   {x_\psi^*}+1\right)$,    
   $\Lambda_{(0,1,2),1}=2 m\sqrt{1+{x_\psi^*}^2}-2 n
   {x_\psi^*}+3$,   
   $\Lambda_{(1,0,2),2}=2 m\sqrt{1+{x_\psi^*}^2}-2 n
   {x_\psi^*}+1$, $\Lambda_{(0,3,0),2}=2$,  $\Lambda_{(0,1,2),2}=2 \left(m\sqrt{1+{x_\psi^*}^2}-n
   {x_\psi^*}+1\right)$, $\Lambda_{(2,0,1),3}=0$, $\Lambda_{(0,2,1),3}=2$,  $\Lambda_{(0,0,3),3}=2 \left(m\sqrt{1+{x_\psi^*}^2}-n
   {x_\psi^*}+1\right)$, are the eigenvalues of the operator linear operator
$${\bf L}^{(3)}_{\bf J}: H^3\rightarrow H^3$$ associated to the
eigenvectors ${\bf x}^{{\bf m}}{\bf e}_i.$
The associated eigenvectors form a basis
of $H^3$ (the space of vector fields with polynomial components of
third degree) because the eigenvalues of ${\bf J}$ are different.

The transformation (\ref{transfz}) is the required by theorem
(\ref{NFTheorem}). By using this theorem we prove the existence of
the required constant $c_3.$ To find its we must to calculate
$$-{\bf L}^{(3)}_{\bf J} {\bf h}_3({\bf z})+\tilde{{\bf X}}_3({\bf z})=\tilde{X}_{(2,0,1),3} z_1^3 {\bf e}_3$$
where ${\bf e}_i,\; i=1,2,3,$ is the canonical basis in
$\mathbb{R}^n.$ Then,
$$c_3=\frac{m}{2 \sqrt{1+{x_\psi^*}^2}}-\frac{n {x_\psi^*}}{2
   \left({x_\psi^*}^2+1\right)}.$$

Observe that the transformation ${\bf h}_3$ does not affect the
value of the coefficient of the resonant term of order $r=2.$
Then, the result of the proposition follows. $\blacksquare$

Using the normal form \eqref{2_17}, \eqref{2_18}, \eqref{2_19} one can calculate approximated invariant manifolds of the origin \cite{Leon:2008aq}. 

If  $\lambda_3^->0$, 
the unstable manifold of the origin is, up to order ${O}(|{\bf
   z}|^4)$, is 
$W^u_{{loc}}({\mathbf 0})=\{\left(z_1,z_2,z_3\right) \in
\mathbb{R}^3: z_1=0, z_2^2+z_3^2<\delta^2\}$ where $\delta$ is a
real value small enough. Therefore, the dynamics of
(\ref{2_17}-\ref{2_19}) restricted to the unstable manifold, is
given, up to order ${\bf O}(|{\bf
   z}|^4),$ by
$z_1\equiv 0,\, z_2(\tau )=e^{\tau } z_{20},\, z_2(\tau
)=e^{\lambda_3^-\tau } z_{30},$ where
$z_{20}^2+z_{30}^2<\delta^2.$ This means that $\lim_{\tau \to
-\infty } \left(z_1(\tau ),
   \, z_2(\tau ),\, z_3(\tau
   )\right)=\left(0,0,0\right).$ Then, the origin is the past attractor
for an open set of orbits of (\ref{2_17}-\ref{2_19}).

For $\lambda_3^->0$, the center manifold of the origin is, up to order ${O}(|{\bf
   z}|^4)$,
$W^c_{{loc}}({\mathbf 0})=\{\left(z_1,z_2,z_3\right) \in
\mathbb{R}^3: z_2=z_3=0, |z_1|<\delta\}$ where $\delta$ is a real
value small enough. Or
as a graph in the original variables defined by
\begin{align} & x_1\equiv x_{\phi
}+\sqrt{{x_\psi^*}^2+1}=-\frac{z_1\left(z_1+2
{x_\psi^*}\right)}{2
\sqrt{{x_\psi^*}^2+1}}+O\left({|z_1|^4}\right)\label{2_45},\\
& x_2\equiv x_{\psi}-{x_\psi^*}=\frac{{x_\psi^*}
z_1^2}{2 {x_\psi^*}^2+2}+z_1+O\left({|z_1|^4}\right)\label{2_46},\\
& x_3\equiv y=O\left({|z_1|^4}\right)\label{2_47}. \end{align}

By taking the inverse, up to fourth order, of (\ref{2_46}) we have
the expression for $z_1$:

\be z_1=\frac{{x_\psi^*}^2 x_2^3}{2
   \left({x_\psi^*}^2+1\right)^2}-\frac{{x_\psi^*} x_2^2}
   {2 \left({x_\psi^*}^2+1\right)}+x_2+O\left({|x_2|^4}\right)\label{2_48}.\ee
Substituting (\ref{2_48}) in (\ref{2_45}-\ref{2_47}) we have that
the center manifold of the origin is given by the graph:
$\{(x_1,x_2,x_3)\in\mathbb{R}^3: x_1=\frac{{x_\psi^*} x_2^3}{2
   \left({x_\psi^*}^2+1\right){}^{5/2}}-\frac{x_2^2}{2
   \left({x_\psi^*}^2+1\right){}^{3/2}}-
   \frac{{x_\psi^*}
   x_2}{\sqrt{{x_\psi^*}^2+1}}+O\left({|x_2|^4}\right), x_3=O\left({|x_2|^4}\right),
   |x_2|<\delta\}$ for $\delta>0$ small enough.

\subsection{Compact variables formulation}
\label{CompactNewflat}

The Friedman equation for $k=0$ can be written as
\begin{equation} H^2 +\sfrac16\dot\psi^2 +\sfrac23 V(\phi,\psi)= \sfrac16 \dot\phi^2 +V(\phi,\psi)+\sfrac13\rho
\end{equation}

It is clear that when  $H^2 +\sfrac16\dot\psi^2 +\sfrac23 V(\phi,\psi)=0$ we obtain for nonnegative potential the trivial solution. Therefore, we consider $H^2 +\sfrac16\dot\psi^2 +\sfrac23 V(\phi,\psi)> 0$. 

We introduce the new formulation 
\begin{align} & X_\phi=\frac{\dot\phi}{\sqrt{6} \bar{D}}, \; 
X_\psi=\frac{\dot\psi}{\sqrt{6} \bar{D}},  \; Y=\frac{\sqrt{V}}{\bar{D}}, \nonumber \\
& \bar{\Omega}=\frac{\rho}{3 \bar{D}^2}, \; \bar{H}= \frac{H}{\bar{D}}, \label{varsXYZO}
\end{align}
where $\bar{D}:=\sqrt{H^2 +\sfrac16\dot\psi^2 +\sfrac23 V(\phi,\psi)}$.
They satisfy 
\begin{align}
X_\phi^2+ Y^2+ 
\bar{\Omega}=1, \quad \bar{H}^2+ X_\psi^2+\frac{2}{3} Y^2=1\label{SignbarH}
\end{align}
That is, we have five variables and two constraints. \\

Therefore, we can study the reduced system
 \begin{align}
& X_\phi'=\frac{1}{3} Y^2 \left(2 m X_{\phi}^2+m+3 n X_{\psi} X_{\phi}\right) \nonumber \\
& +\frac{1}{6} X_{\phi}\left(X_{\phi}^2-Y^2-1\right) \varepsilon  \sqrt{9-9 X_{\psi} ^2-6 Y^2},\label{EQ58}
\\
&  X_\psi'=\frac{1}{3} Y^2 \left(2 m X_{\psi}  X_{\phi}+n \left(3 X_{\psi} ^2-1\right)\right) \nonumber \\
& +\frac{1}{6} X_{\psi} \left(X_{\phi}^2-Y^2-1\right)
   \varepsilon  \sqrt{9-9 X_{\psi} ^2-6 Y^2},\label{EQ59}
\\
   & Y'=\frac{1}{3} m X_{\phi} Y \left(2 Y^2-3\right)+n X_{\psi}  Y
   \left(Y^2-1\right)\nonumber \\ &+\frac{1}{6} Y\left(X_{\phi}^2-Y^2+1\right) \varepsilon  \sqrt{9-9 X_{\psi} ^2-6 Y^2},\label{EQ60}
 \end{align}
defined in the compact phase space
\begin{equation}
\{(X_\phi, X_\psi, Y)\in\mathbb{R}^3: X_\phi^2+ Y^2\leq 1, X_\psi^2+\frac{2}{3} Y^2\leq 1\}
\end{equation}
where the prime means derivative 
\begin{equation}
'\equiv \frac{d}{d\bar{\tau}}=\frac{1}{3 \bar{D}} \frac{d}{dt}\label{timeXYO}.
\end{equation}
and $\varepsilon=\pm1$ appears due to \eqref{SignbarH} cannot be solved globally for $\bar{H}$, but $\bar{H}=\pm \frac{\sqrt{3-3 X_{\psi}^2-2 Y^2}}{\sqrt{3}}$, and it corresponds to the sign of $H$, where $\varepsilon=+1$ corresponds to ever expanding universe and $\varepsilon=-1$ corresponds to ever contracting universes.

\begin{figure*}[ht]
\centering
\subfigure[\;\;$m=0.5, n=0.6, \gamma=1$.]{\includegraphics[width=0.3\textwidth]{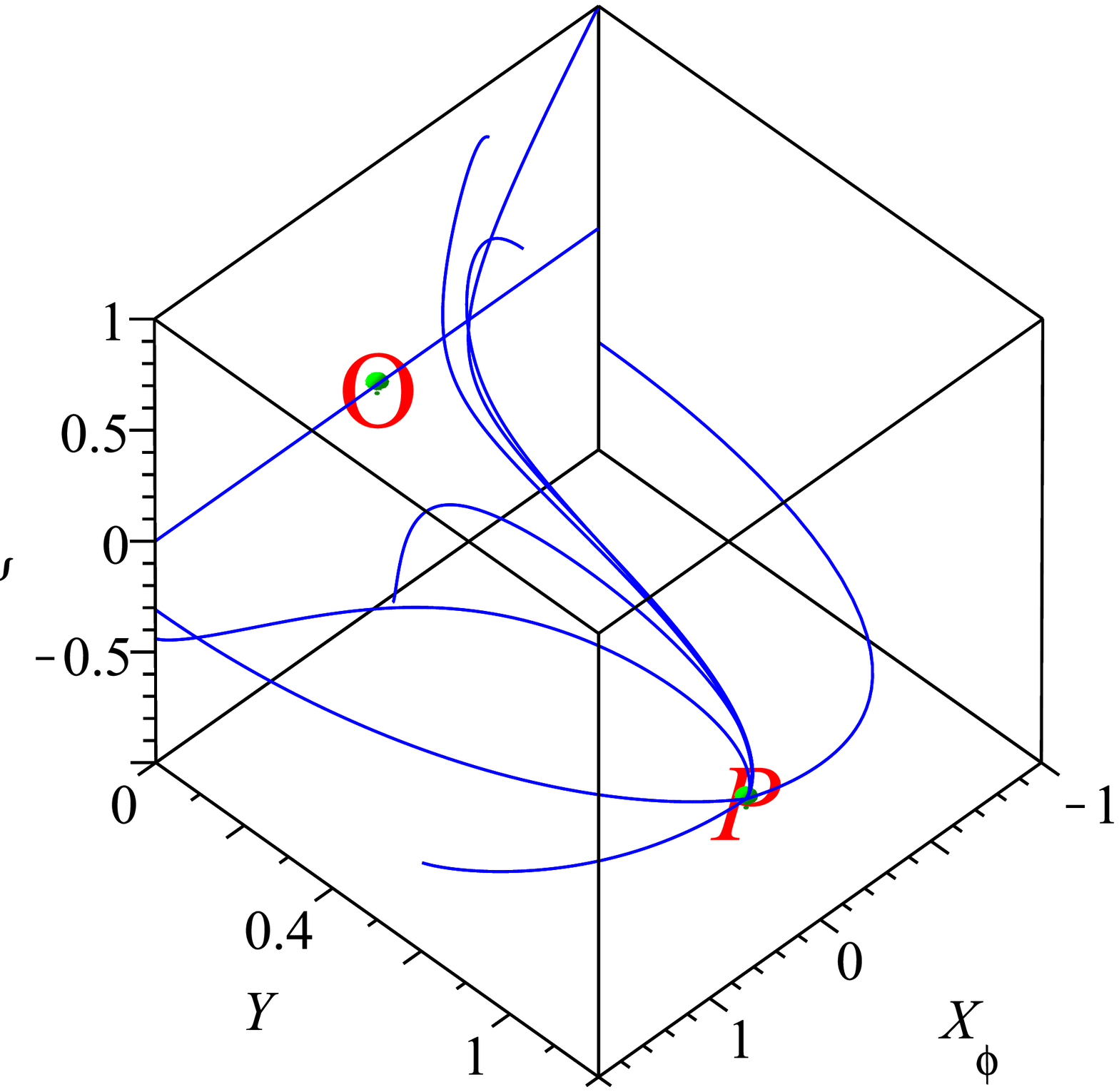}}
\subfigure[\;\;$m=0.75, n=0.05, \gamma=1$.]{\includegraphics[width=0.3\textwidth]{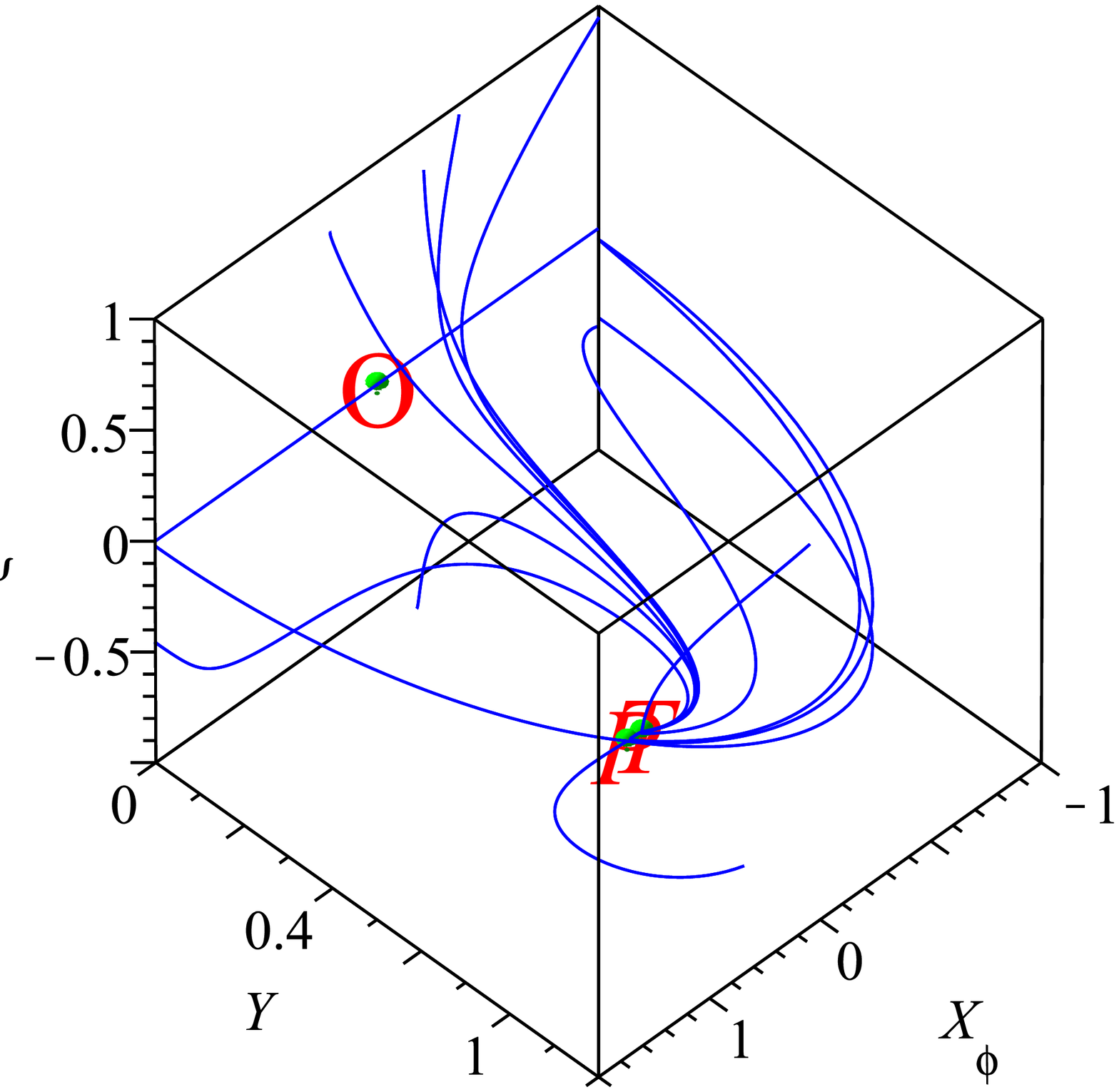}}
\centering
\subfigure[\;\;$m=0.9, n=0.4, \gamma=1$.]{\includegraphics[width=0.3\textwidth]{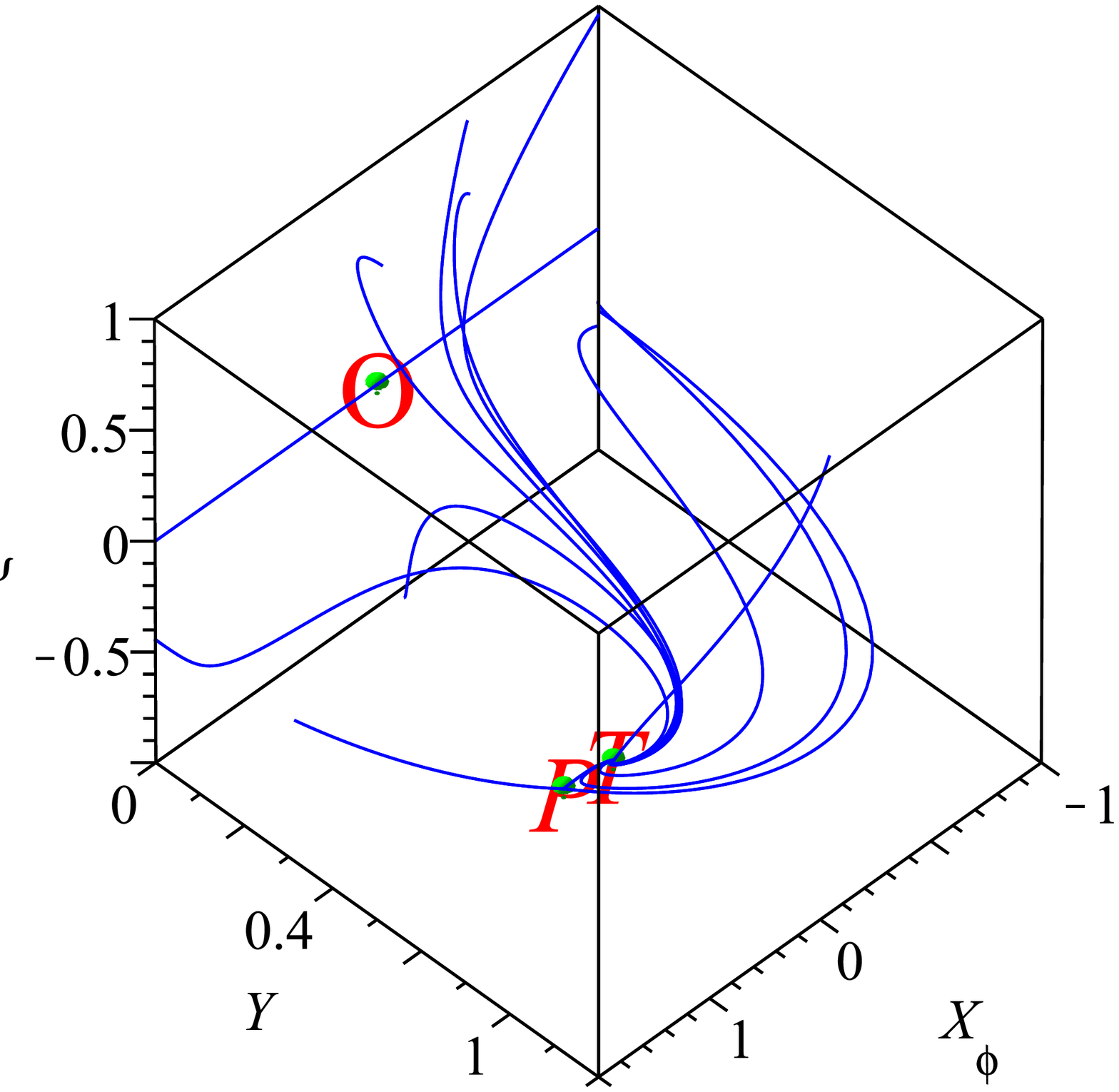}}
\centering
\subfigure[\;\;$m=2.0, n=0.5, \gamma=1$.]{\includegraphics[width=0.3\textwidth]{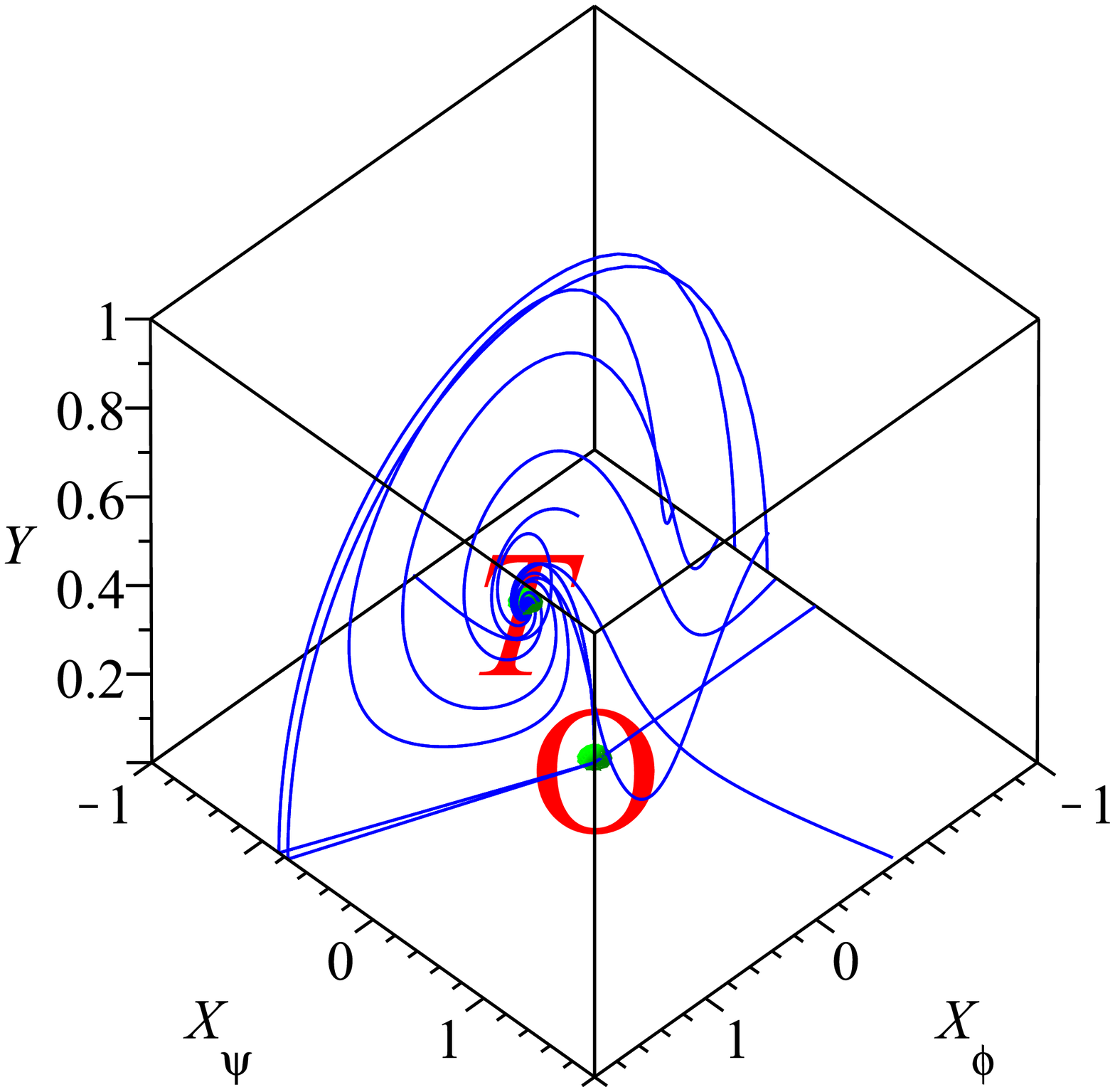}}
\subfigure[\;\;$m=0.3, n=0.5, \gamma=1$.]{\includegraphics[width=0.3\textwidth]{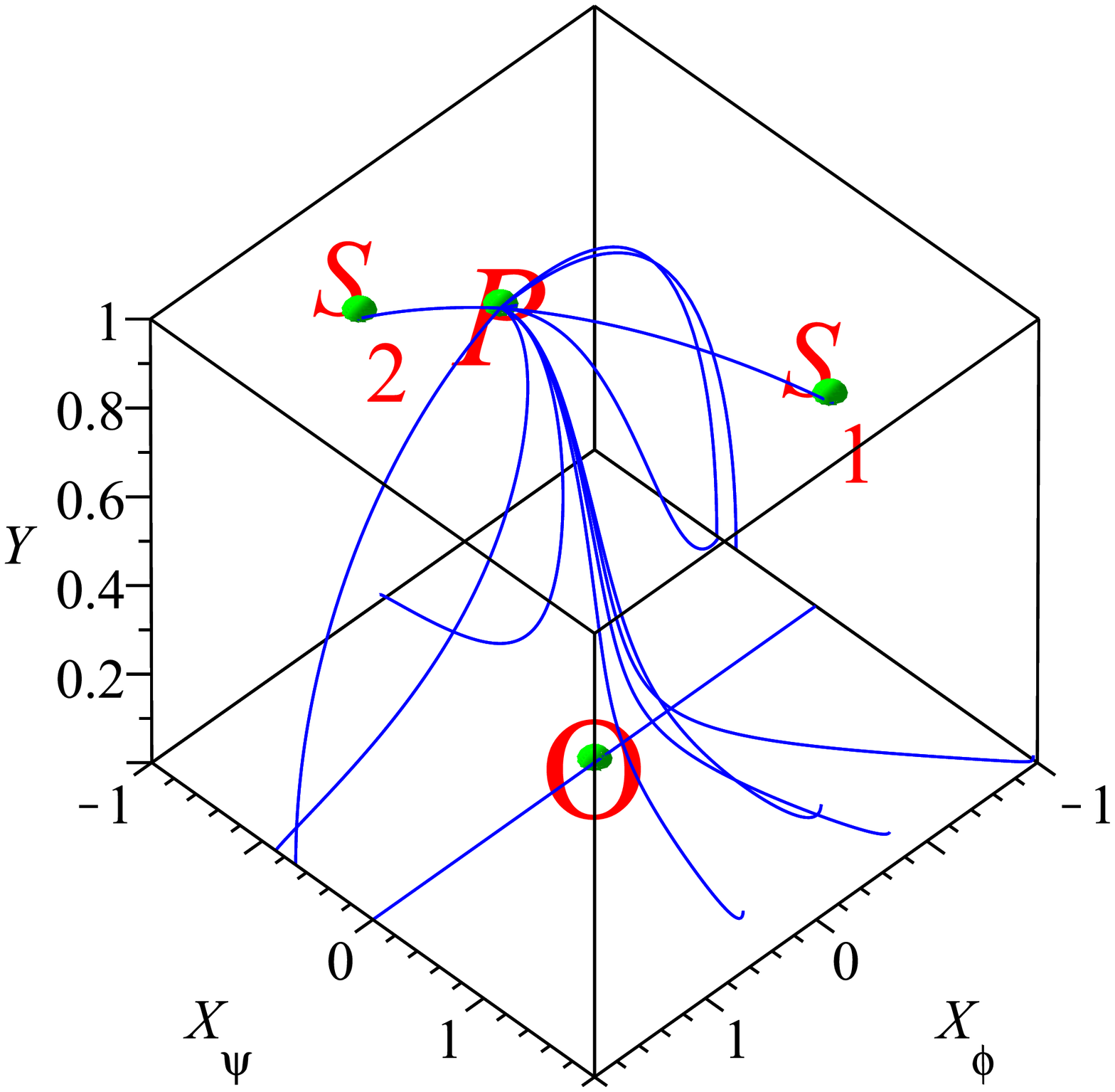}}
\caption{\label{FIG3} Some orbits in the phase plane of the system \eqref{EQ58}, \eqref{EQ59}, \eqref{EQ60} for the five choices of the parameters and $\varepsilon=+1$.}
\end{figure*}

The equilibrium points/lines of the system \eqref{EQ58}, \eqref{EQ59}, \eqref{EQ60} (for a fixed sign of $\varepsilon$) have coordinates $(X_\phi, X_\psi, Y)$:
\begin{enumerate} 
\item $O_{\pm}:=(0,0,0)$. 

For $\varepsilon=+1$ we obtain a representation of the point $O$ referred in table \ref{table1}.  

\item $C_\pm:=(\pm 1, X_\psi^{\star}, 0)$ referred in table \ref{table1}.

\item $P_\pm:=\Big(\frac{m \varepsilon }{\sqrt{-2 m^2+3 n^2+3}},-\frac{n \varepsilon }{\sqrt{-2 m^2+3 n^2+3}},$\\
$\sqrt{3} \sqrt{\frac{m^2-n^2- 1}{2 m^2-3 n^2-3}}\Big)$. The point $P$ referred in table \ref{table1} is recovered by setting $\varepsilon=+1$.

\item  $T_\pm:=\Big(\frac{m \varepsilon }{\sqrt{4 \left(m^2-n^2\right)^2+2 m^2-n^2}},-\frac{n \varepsilon }{\sqrt{4 \left(m^2-n^2\right)^2+2 m^2-n^2}},$\\$\frac{\sqrt{3} \sqrt{m^2-n^2}}{\sqrt{4
   \left(m^2-n^2\right)^2+2 m^2-n^2}}\Big)$. The point $T$ referred in table \ref{table1} is recovered by setting $\varepsilon=+1$.

\item 
 {\small
 \begin{enumerate}
 \item $S_1:=\left(-\frac{m}{\sqrt{3 n^2-2 m^2}},  \frac{n}{\sqrt{3 n^2-2 m^2}},\sqrt{3} \sqrt{\frac{m^2-n^2}{2 m^2-3 n^2}}\right)$. 
 \item $S_2:=\left(\frac{m}{\sqrt{3 n^2-2 m^2}}, -\frac{n}{\sqrt{3 n^2-2 m^2}},\sqrt{3} \sqrt{\frac{m^2-n^2}{2 m^2-3 n^2}}\right)$.  
 The static solutions $S_1$ and $S_2$ exists for  
 $n>0, -n\leq m\leq n$.

\item ${}_{\pm} S_3:=(X_\phi^{\star}, \pm 1, 0)$. The lines ${}_{\pm}S_3$ exists for $-1\leq X_\phi^{\star}\leq 1$.  
 
 \end{enumerate}
 }
 located at the curve $S:= 3 X_{\psi}^2+2 Y^2-3=0$. They represent static universes $\bar{H}=H=0$. 
\end{enumerate}

In figure \ref{FIG3} (a), (b), (c), and (d) and (e) are presented some orbits in the phase plane of the system \eqref{EQ58}, \eqref{EQ59}, and \eqref{EQ60} for $\varepsilon=+1$, that resembles the results found in \cite{31}.
The static universe configurations cannot by found
in the framework of \cite{31}. This illustrates the applicability
of this formulation in compact variables.

\section{Alternative normalization}  
\label{crossing}

Due the negative sign of the kinetic energy of the phantom
field, it is possible that H can be zero in a finite time.
If so, the variables in \cite{31} are not well defined as $H\rightarrow 0$. 
They can diverge in a finite parameter time, whence the
resulting system is not a dynamical system. To avoid this
difficulty we will use an alternative set of dynamical variables, albeit non compact, similar to those introduced in \cite{Coley:2003mj} section III.E, instead of the ones used in \cite{31}, for investigating negative (and zero curvature models). For investigating positive curvature models we shall make use of similar variables to those defined in \cite{Coley:2003mj} section VI.A. 

\subsection{Negative and zero curvature models} 

In this section we extent the analysis presented in \cite{Leon:2009ce}, since now it is included the possibility of the crossing through $H=0$. 

Let us introduce the following set of normalized variables: $(D,\,U_\phi,\,U_\psi,\,W,\,\Omega)$, given by
\begin{eqnarray}
&D= \frac{3 H}{\sqrt{1+9 H^2}},\; U_\phi=\sqrt{\frac{3}{2}}\frac{\dot\phi}{\sqrt{1+9 H^2}},\;U_\vphi=\sqrt{\frac{3}{2}}\frac{\dot\psi}{\sqrt{1+9 H^2}},\nonumber\\
& W=\frac{\sqrt 3 V}{\sqrt{1+9 H^2}},\,\Omega= \frac{3 \rho}{1+9 H^2}.\label{vars}
\end{eqnarray}  This choice of phase-space variables renders the Friedmann equation as
\begin{eqnarray}
&&D^2-\left(U_\phi^2-U_\psi^2+W^2+\Omega\right)= \Omega_k D^2\geq 0,\nonumber \\
&&\Omega_k=-\frac{k}{a^2 H^2},\; k=-1,\,0.
\label{ct}
\end{eqnarray}

By definition $-1\leq D\leq 1.$ Then, 

\begin{equation}
0\leq U_\phi^2-U_\psi^2+W^2+\Omega\leq D^2\leq 1 \label{constraint}.
\end{equation}
From the requirement  of the non-negativeness of the energy densities of quintom and matter, follows $\Omega=\Omega_m D^2\geq 0$ and $\Omega_{de} D^2\equiv U_\phi^2-U_\psi^2+W^2\geq 0$. $\Omega_m,\, \Omega_{de},\, \Omega_k$ denotes the fractional energy densities of dark matter, dark energy and ``curvature''. Then, from (\ref{constraint}) follows that $\Omega$ is bounded as $\Omega_{de}$ is. However, we cannot guaranteed that the variables $U_\phi,\, U_\psi,$ $W$ are bounded. In this sense, the dynamical system constructed from the variables (\ref{vars}) is unbounded (then, non-compact). Hence, we cannot guaranteed that the system admits both future and past attractors. Numerical experience (see for instance \cite{31}) supports the conjecture that past and future attractors can exists depict unboundedness. 

For $D\neq 0$ (i.e., $H\neq 0$) the variables (\ref{vars}) are related with those in \cite{31} via 

\begin{equation}
\left(x_\phi^2,\, x_\psi^2,\,y^2,\,z^2\right)=\left(U_\phi^2,\, U_\psi^2,\,W^2,\,\Omega\right)/D^2.
\end{equation}

The equation of state (EoS) parameter, $w,$ and the deceleration parameter, $q,$ can be stated in terms of our variables as 

\begin{equation}
w =\frac{U_\phi^2-U_\psi^2-W^2}{U_\phi^2-U_\psi^2+W^2},\, q=-1+Q/D^2,
\end{equation} where  \begin{equation} Q=D^2+2\left(U_\phi^2-U_\psi^2\right) -W^2+\frac{1}{2}\left(3\gamma-2\right)\Omega.\label{Q}\end{equation}

The evolution equations for (\ref{vars}) are 

\begin{eqnarray}
&& D' =-\frac{1}{3}\left(1-D^2\right) Q,\nonumber\\
&& U_\phi'=- D\,\left( 1 - \frac{Q}{3} \right) \,U_\phi + m\,W^2,\nonumber\\
&& U_\psi'=-D\,\left( 1 - \frac{Q}{3} \right) \,U_\psi  -n\,W^2,\nonumber\\
&& W'=-\left( - \frac{1}{3}\,D\,Q + m\,U_\phi + n\,U_{\psi}
      \right) \,W, \nonumber\\
&& \Omega'=- D\,\left( -\frac{2}{3}\,Q + \gamma  \right) \,\Omega,\label{PhaseSpaceeqs}
\end{eqnarray} 
where we have defined the new independent variable 

\begin{equation}
'\equiv \frac{d}{d\tau}=\frac{1}{\sqrt{1+9 H^2}} \frac{d}{dt}\label{time}.
\end{equation}

The ODEs (\ref{PhaseSpaceeqs}) define a flow on the state space 

\begin{eqnarray}
\Psi= & \{\left(D, U_\phi, U_\psi, W,\Omega\right)\in\mathbb{R}^5: \nonumber\\
& 0\leq U_\phi^2-U_\psi^2+W^2+\Omega \leq D^2, -1\leq D\leq 1, \nonumber\\
& U_\phi^2-U_\psi^2+W^2\geq 0, W\geq 0,\Omega\geq 0\}.\label{space}
\end{eqnarray}

Observe that the variable $D$ does not decouples from the other variables. Hence, the system is of higher dimension (5-dimensional) that the system we can obtain by using the variables defined in \cite{31} (4-dimensional in the case of negative curvature). 
\begin{rem}
In the case $D=0$ we find from (\ref{constraint}) and from the requirement of non-negativeness of the energy densities of the sources, that $\Omega=0$ and $U_\phi^2-U_\psi^2+W^2=0.$ Then, $Q|_{D=0}=-3 W^2\leq 0\Rightarrow D'|_{D=0}=-\frac{1}{3} Q|_{D=0}= W^2\geq 0.$ From the last inequality we can conclude that the submanifold $D=0$ is not invariant for models with $W>0,$ but acts as a membrane. For massless scalar field (MSF) cosmologies, the submanifold $D=0$ is invariant and is equivalent to the set $\Omega=0,U_\phi^2-U_\psi^2=0$.
\end{rem}
The dynamical system (\ref{PhaseSpaceeqs}) is invariant  under the transformation of coordinates 

\begin{equation}
\left(\tau, D, U_\phi, U_\psi,W,\Omega\right)\rightarrow \left(-\tau, -D, -U_\phi, -U_\psi,W,\Omega\right)\label{discrete}.
\end{equation}
Thus, it is sufficient to discuss the behavior in one part of the phase space. The dynamics in the other part being obtained via the transformation (\ref{discrete}).

\begin{rem}
Let be defined in the interior of the phase space the function 

\begin{align}
& M=\frac{{\left( n\,U_\phi + m\,U_\psi \right) }^2\,{\Omega }^3}
    {\left(1 - D^2 \right) \,{\left( D^2 - {U_\phi}^2 + {U_\psi}^2 - 
          W^2 - \Omega  \right) }^3},\nonumber \\
          &M'=-3\gamma D M. \label{monotonic1}
\end{align}
This function is a monotonic function in the regions $0<D<1$ and $-1<D<0$ for $\Omega>0$ and $n\,U_\phi + m\,U_\psi\neq 0.$  From the expression of $M$ we can see immediately that either $D^2\rightarrow 1$ or $\Omega\rightarrow 0,$ or $n\,U_\phi + m\,U_\psi \rightarrow 0$ or $\left|n\,U_\phi + m\,U_\psi\right| \rightarrow +\infty$ (which means that $U_\phi $ or $U_\psi$ or both diverge) or $\Omega_k\rightarrow 0$ asymptotically. Particularly, $M$ is a monotonic decreasing function for models with $0<D<1$ (i.e., for ever expanding models) and takes values in the interval $(0,+\infty).$ Since $M\rightarrow 0$ as $\Omega\rightarrow 0$ or as $n\,U_\phi + m\,U_\psi\rightarrow 0$ (since its denominator is always bounded), then, the future asymptotic state of the system for ever expanding models corresponds to models with $\Omega=0$ or to models where $n\,U_\phi + m\,U_\psi=0.$ Since $M\rightarrow +\infty$ as $\left|n\,U_\phi + m\,U_\psi\right| \rightarrow +\infty$ or as $D \rightarrow 1$ or as $\Omega_k\rightarrow 0$, then the past asymptotic state of the system for ever expanding models corresponds to models where at least one of the variables $U_\phi$ or $U_\psi$ diverges, or to models with $H$ large, or to flat models. 
\end{rem}
The asymptotic dynamics for contracting models can determined from the asymptotic dynamics of ever expanding models by means of the coordinate transformation (\ref{discrete}). 

\begin{rem}
The existence of the monotonic function \eqref{monotonic1} in our state space rules out any periodic, recurrent, or homoclinic orbit in the interior of the phase space. Therefore, some global results can be found from the linear (local) analysis of equilibrium points.
\end{rem}

In the table \ref{T1} it is displayed the conditions for existence and the eigenvalues of the linearized system around each equilibrium point of the system (\ref{PhaseSpaceeqs}). In table \ref{T2} we give the cosmological parameters associated to the equilibrium points. In the following we will characterize them. 

\begin{table*}[t!]
\begin{tabular}{|c|c|c|c|}
\hline
Label & $(D, U_\phi,U_\psi,W,\Omega)$& Existence& Eigenvalues \\
\hline
$V_1$ & $(0,-U_\psi^\star,U_\psi^\star,0,0)$ & All $m$ and $n$ & $0[\times 4],(m-n)U_\psi^\star$ \\[0.2cm]
$V_2$  & $(0,U_\psi^\star,U_\psi^\star,0,0)$ & All $m$ and $n$ & $0[\times 4],-(m+n)U_\psi^\star$ \\[0.2cm]
${}_{\pm}K_{\pm}$ & $(\epsilon,\pm\sqrt{1+{U_\psi^\star}^2},U_\psi^\star,0,0)$ & All $m$ and $n$ & $2\epsilon, \frac{4}{3}\epsilon,0,\epsilon-n U_\psi-m U_\phi,\epsilon(2-\gamma)$ \\[0.2cm]
${}_{\pm}M$ & $(\epsilon,0,0,0,0)$ & All $m$ and $n$ & $\frac{2}{3},-\frac{2}{3},-\frac{2}{3}\epsilon,\frac{1}{3}\epsilon,-\left(\gamma-\frac{2}{3}\right)\epsilon$ \\[0.2cm]
${}_{\pm}F$ & $(\epsilon,0,0,0,1)$ & All $m$ and $n$ & $\gamma\epsilon,\frac{\gamma}{2}\epsilon,\left(\frac{\gamma}{2}-1\right)\epsilon[\times 2],\left(\gamma-\frac{2}{3}\right)\epsilon$ \\[0.4cm]
${}_\pm SF$ & $(\epsilon,m\epsilon,-n\epsilon,\sqrt{1-\delta},0)$ & $\delta<1$ & $2\delta\epsilon,-2\left(\frac{1}{3}-\delta\right)\epsilon,-\left(1-\delta\right)\epsilon[\times 2],\left(2\delta-\gamma\right)\epsilon$ \\[0.2cm]
${}_\pm CS$  & $(\epsilon,\frac{m \epsilon}{3\delta},-\frac{n \epsilon}{3\delta},\frac{\sqrt{2}}{3\sqrt{\delta}},0)$ & $\delta>\frac{1}{3}$& $\frac{2}{3}\epsilon,-\frac{2}{3}\epsilon,\left(\frac{2}{3}-\gamma\right)\epsilon,-\frac{1}{3}\left(1\pm\sqrt{\frac{4}{3\delta}-3}\right)\epsilon$ \\[0.2cm]
${}_{\pm}MS$ & $(\epsilon,\frac{m\gamma\epsilon}{2\delta},-\frac{n\gamma\epsilon}{2\delta},\frac{\sqrt{(2-\gamma)\gamma}}{2\sqrt{\delta}},\sqrt{1-\frac{\gamma}{2\delta}})$ & $\delta>\frac{\gamma}{2}$  & $\gamma\epsilon,-\left(\frac{2}{3}-\gamma\right)\epsilon,\left(-1+\frac{\gamma}{2}\right)\epsilon, \lambda^\pm\epsilon$ \\[0.4cm]
\hline
\end{tabular}
\caption{\label{T1}equilibrium points of quintom model with $k=0,-1$. We use the notations $\delta={m}^2-{n}^2$ and $\ds \lambda^\pm= -\frac{1}{4}\left(2-\gamma\pm \sqrt{(2-\gamma)\left(2-9\gamma+\frac{4\gamma^2}{\delta}\right)}\right).$ $[\times s]$ means multiplicity s. The subscripts on the label have the following significance. The left subscript gives the sign of $D$ (denoted by $\epsilon=\pm 1$) and indicates whether the corresponding model is expanding ($+$) or contracting ($-$). The right subscript gives the sign of $U_\phi$ (i.e., the sign of $\dot \phi$) and is denoted by the sign $\pm$. When the flow is restricted to the invariant sets $D=\pm 1,$ the eigenvalues associated to the equilibrium points ${}_{\pm}{M},$ ${}_{\pm}{F},$ ${}_\pm {SF}$ and ${}_\pm {MS}$ an to the equilibrium sets ${}_{\pm}\hat{K}_{\pm},$ are, in each case, the same as those displayed, but the first from the left.}
\end{table*}

The sets of equilibrium points $V_{1,2}$ and ${}_{\pm}K_{\pm}$ and the isolated equilibrium points ${}_{\pm}M$ are located in the invariant set of MSF cosmologies without matter. The isolated equilibrium point ${}_{\pm}F$ is located in the invariant set of MSF cosmologies with matter.  

$V_{1,2}$ are degenerated equilibrium points. They represent flat vacuum static solutions $\left(\Omega_{de}\rightarrow 0, \Omega_{m}\rightarrow 0,\Omega_{k}\rightarrow 0\right).$ 

The arcs ${}_{\pm}K_{\pm}$ denotes cosmological models dominated by dark energy ($\Omega_{de}\rightarrow 1$) particularly by its kinetic energy. The DE mimics a stiff fluid solution. The stable (resp., unstable) manifold of  ${}_{-}K_{\pm}$ (resp., ${}_{+}K_{\pm}$ ) is at least 3-dimensional in the full phase space. 

The equilibrium points ${}_{\pm}M$ denotes the Milne's universe. They are non hyperbolic for $\gamma=\frac{2}{3}.$ The equilibrium points ${}_{\pm}F$ represent the flat FRW universe. They are non-hyperbolic if $\gamma=\frac{2}{3}$ or $\gamma=2.$ For $\gamma>\frac{2}{3},$ both ${}_{\pm}M$ and  ${}_{\pm}P$ acts locally as saddles. The unstable (resp., stable) manifold of ${}_{-}M$ (resp. ${}_{+}M$) is 3-dimensional in the full phase space. The equilibrium point ${}_{+}M$ (resp. ${}_{-}M$) is unstable (resp. stable) to perturbations in $D$ and in $W.$ The stable (resp., stable) manifold of ${}_{-}F$ (resp. ${}_{+}F$) is 3-dimensional. Whereas, the unstable (resp. stable) manifold ${}_{-}F$ (resp. ${}_{+}F$) is 2-dimensional in the full phase space. If we restrict the dynamics to the invariant sets with $D^2=1$ the dimensionality of all the above invariant sets reduces in one unity.

The isolated equilibrium points ${}_\pm SF$ and ${}_\pm CS$ denotes scalar field dominated solutions and curvature scaling solutions, respectively. They are located in the invariant set $W>0,\Omega=0.$ The equilibrium points ${}_\pm MS$  (belonging to the invariant set $W>0,\Omega>0$) represents flat matter scaling solutions.  Observe that the equilibrium points ${}_\pm MS$ and  ${}_\pm SF$ coincide as $\delta\rightarrow \frac{\gamma}{2}^+$ and ${}_\pm CS$ and  ${}_\pm SF$ coincide as $\delta\rightarrow\frac{1}{3}^+.$ Additionally, ${}_+ SF$ (resp., ${}_- SF$) coincides with a point in the arc ${}_+K_+$ (resp., ${}_-K_-$) as $\delta\rightarrow 1^-.$ These values of $\delta$ where equilibrium points coincide corresponds to bifurcations. 

Due the symmetry (\ref{discrete})  it is sufficient to characterize the behavior of the equilibrium points ${}_+ SF,$  ${}_+ CS,$ and ${}_+ MS.$ The dynamical character of their  counterparts in the ``negative'' branch  ${}_- SF,$  ${}_- CS,$ and ${}_- MS$ is determined by the symmetry (\ref{discrete}).

\begin{table}[t!]
\begin{tabular}{|c|c|c|c|}
\hline
Label &  $w$& $q$&  $\Omega_m,\Omega_{de},\Omega_k$
\\\hline
$V_{1,2}$ & indeterminate & indeterminate & indeterminate \\[0.2cm]
${}_{\pm}K_{\pm}$ & $1$ & $2$ & $0,1,0$ \\[0.2cm]
${}_{\pm}M$ & indeterminate & $0$ & $0,0,1$ \\[0.2cm]
${}_{\pm}F$ & indeterminate & $-1+\frac{3\gamma}{2}$ & $1,0,0$  \\[0.2cm]
${}_\pm SF$ & $-1+2\delta$ & $-1+3\delta$ & $0,1,0$ \\[0.2cm]
${}_\pm CS$  & $-\frac{1}{3}$ & $0$& $0,\frac{1}{3\delta},1-\frac{1}{3\delta}$ \\[0.2cm]
${}_{\pm}MS$ & $-1+\gamma$ & $-1+\frac{3\gamma}{2}$  & $1-\frac{\gamma}{2\delta},\frac{\gamma}{2\delta},0$ \\[0.4cm]
\hline
\end{tabular}
\caption{\label{T2}Cosmological parameters associated to the equilibrium points.}
\end{table}

The equilibrium point ${}_+SF$ is non hyperbolic in the full phase space if either $\delta=\sfrac{1}{3}$ or $\delta=\sfrac{\gamma}{2}$ or $\delta=0.$ However, if we restrict our attention to the flow in the invariant set $D=1,$ the eigenvalue $2\delta$ plays no role in the dynamics (it appears due the ``extra'' dimension, $D$). In fact, if $\delta=0,$ the eigenvalues of the system restricted to $D=1$ are $(-2/3,-1[\times 2],-\gamma$). Then,  ${}_+SF$ represents a de Sitter solution, and it is the attractor in the invariant set $D=1.$ If $\delta<0$ and $\sfrac{2}{3}<\gamma<2$ then, ${}_+SF$ is the future attractor (in the full phase space) for (ever) expanding models and it represents a phantom field. Using the symmetry (\ref{discrete}) we find that, if $\delta<0$ and $\sfrac{2}{3}<\gamma<2$ then, ${}_-SF$ is the past attractor for contracting models and it represents a phantom field. If $0<\delta<\sfrac{1}{3}$ and $\sfrac{2}{3}<\gamma<2,$ then ${}_+SF$ represents a solution dominated by a quintessence field, and it is the future attractor for the flow restricted to the invariant set $D=1$. If $0<\delta<\sfrac{1}{3}$ and $\sfrac{2}{3}<\gamma<2,$ then ${}_-SF$ represents a solution dominated by a quintessence field, and it is the past attractor for the flow restricted to the invariant set $D=-1$.

The equilibrium point ${}_+CS$ represents a solution with negative curvature. It is non hyperbolic if $\delta=\sfrac{1}{3}.$ It is a saddle point for the full phase space, since it have always at least a positive eigenvalue. It is never a source (all the eigenvalues are not simultaneously with positive real parts) for the class of ever expanding models. By using the symmetry argument we find that ${}_-CS$ is never a global attractor (or sink) for contracting models. But, if we restrict our attention to the flow in the invariant set $D=1$ we find that in this case ${}_+CS$ is the future attractor provided $\delta>\sfrac{1}{3}$ and $\sfrac{2}{3}<\gamma<2.$ In other words, one can say that if $\sfrac{2}{3}<\gamma<2$ the spatial curvature destabilizes ${}_+SF$ at the bifurcation value $\delta=\sfrac{1}{3}$ and that the stability is transferred to the equilibrium point ${}_+CS$ (of course, this argument is valid only in the invariant set $D=1$). 

The equilibrium point ${}_+MS$ represents a cosmological solution where DE tracks DM (it means that the DE EoS parameter scales as the DM EoS parameter). It is non hyperbolic for $\gamma={2}/{3}$ or $\gamma=2.$ It is in general a saddle point. However, when the dynamics is restricted to the invariant set $D=1$ it can be the future attractor (for ever expanding models) provided $0<\gamma<\sfrac{2}{3}$ and $\delta>\sfrac{\gamma}{2}.$ From our hypothesis, this is not a region of physical interest. However, if we do not restrict the value of $\gamma$ to be $>\sfrac{2}{3}$ the above fact can be interpreted as follows. If $0<\gamma<\sfrac{2}{3}$ the matter destabilizes ${}_+SF$ at the bifurcation value $\delta=\sfrac{\gamma}{2}$ and that the stability is transferred to the equilibrium point ${}_+MS.$ Otherwise, whenever $0<\gamma<\sfrac{2}{3}$ and $\delta<\sfrac{\gamma}{2}$ the late time attractor in the invariant set $D=1$ is the equilibrium point ${}_+SF.$

These equilibrium points are such that $n\,U_\phi + m\,U_\psi=0.$ Hence, by the structure of the function (\ref{monotonic1}), for ever expanding models (i.e., for $0<D<1$), the late time behavior is determined by the local behavior of either ${}_+ SF$ or ${}_+ CS$ or ${}_+ MS.$ By making use of the symmetry (\ref{discrete}) it is possible to determine the early time behavior of contracting models (i.e, for $-1<D<0$) from the local behavior of either ${}_- SF$ or ${}_- CS$ or ${}_- MS.$ 
Combining the above linear analysis, with the information we have using monotonic functions we have the following 
\begin{prop}
The past and future attractors of the quintom model with $k=0,-1$ are as follows: 
\begin{enumerate}
\item For $D=-1$: 
\begin{enumerate}
\item The past attractors are: 
\begin{enumerate} 
\item ${}_-SF$ if $\frac{2}{3}\leq\gamma\leq 2,\,0\leq\delta<\sfrac{1}{3}$ or \\
$0\leq\gamma\leq \sfrac{2}{3},\, 0\leq\delta<\sfrac{\gamma}{2}$.
\item 
${}_-CS$ if $\sfrac{2}{3}<\gamma<2,\,\delta>\sfrac{1}{3}$. 
\item ${}_-MS$  if $0<\gamma<\sfrac{2}{3},\,\delta>\sfrac{\gamma}{2}$.
\end{enumerate}

\item The future attractor is  ${}_-K_\pm\,\text{if}\;\,n U_\psi^\star\pm m\sqrt{1+{U_\psi^\star}^2}>-1$. 
\end{enumerate}

\item For 
$-1<D<0$ 
\begin{enumerate}
\item The past attractor is $ {}_-SF$ if $0\leq\gamma\leq 2,\,\delta<0$.
\item The future attractor  lies on  ${}_-K_\pm\,\text{if}\;\,n U_\psi^\star\pm m\sqrt{1+{U_\psi^\star}^2}>-1$. 
\end{enumerate}
\item For 
$0<D<1$ 
\begin{enumerate}
\item The past attractor is  ${}_+K_\pm$ if $n U_\psi^\star\pm m\sqrt{1+{U_\psi^\star}^2}<1$. 
\item The future attractor is $ 
{}_+SF$ if $0\leq\gamma\leq 2,\,\delta<0$.
 \end{enumerate}
 
 \item For 
$D=1$ 

\begin{enumerate}
\item The past attractor is  ${}_+K_\pm$ if $n U_\psi^\star\pm m\sqrt{1+{U_\psi^\star}^2}<1$.

\item The future attractors are
\begin{enumerate}
\item ${}_+SF$ if $\frac{2}{3}\leq\gamma\leq 2,\,0\leq\delta<\sfrac{1}{3}$ or $0\leq\gamma\leq \sfrac{2}{3},\, 0\leq\delta<\sfrac{\gamma}{2}$
\item 
${}_+CS$ if $\sfrac{2}{3}<\gamma<2,\,\delta>\sfrac{1}{3}$.
\item ${}_+MS$ if $0<\gamma<\sfrac{2}{3},\,\delta>\sfrac{\gamma}{2}$.
\end{enumerate}
\end{enumerate}
\end{enumerate}
\end{prop}

\subsection{Positive curvature models ($k=+1$)}

In this section we reproduce the results found by one of us in \cite{Leon:2009ce}. Introducing the normalization factor 
\begin{equation}
\hat{D}= 3\sqrt{H^2+a^{-2}}.
\end{equation}
we have $$\hat{D}\rightarrow 0\Leftrightarrow H\rightarrow 0,\, a\rightarrow +\infty$$ (i.e., at a singularity). This means that it is not possible that $\hat{D}$ vanishes at a finite time. Now, using the normalized variables
\begin{eqnarray}
&Q_0= \frac{3 H}{\hat{D}},\; \hat{U}_\phi=\sqrt{\frac{3}{2}}\frac{\dot\phi}{\hat{D}},\;\hat{U}_\vphi=\sqrt{\frac{3}{2}}\frac{\dot\psi}{\hat{D}},\nonumber\\
& \hat{W}=\frac{\sqrt 3 V}{\hat{D}},\,\Omega= \frac{3 \rho}{\hat{D}}.\label{vars2}
\end{eqnarray}

From the Friedmann equation we find 

\begin{equation}
0\leq \hat{U}_\phi^2-\hat{U}_\psi^2+\hat{W}^2=1-\hat{\Omega}\leq 1\label{Friedmann}
\end{equation} 

and by definition 

\begin{equation}
-1\leq Q_0\leq 1. \label{Q0}
\end{equation}

By the restrictions (\ref{Friedmann}, \ref{Q0}), the state variables are in the state space

\begin{align}
 \hat{\Psi}=&\Big\{\left(Q_0,\hat{U}_\phi,\hat{U}_\psi,\hat{W}\right): 0\leq \hat{U}_\phi^2-\hat{U}_\phi^2+\hat{W}^2\leq 1, \nonumber  \\
  & -1\leq Q_0\leq 1\Big\}.\label{statespace2}
\end{align}
As before, this state space is not compact. 

Let us introduce the time coordinate $$' \equiv \frac{d}{d\hat{\tau}}=\frac{3}{\hat{D}}\frac{d}{d t}.$$

$\hat{D}$ has the evolution equation 

$$\hat{D}'=-3 Q_0\hat{D}\left(\hat{U}_\phi^2-\hat{U}_\psi^2+\frac{\gamma}{2}\hat{\Omega}\right)$$ where 
$$\hat{\Omega}=1-\left(\hat{U}_\phi^2-\hat{U}_\psi^2+\hat{W}\right).$$ This equation decouples from the other evolution equations. Thus, a reduced set of evolution equations is obtained. 
{\small
\begin{align}
& Q_0'= {\left( 1 - Q_0^2 \right) \,\left( 1 - 
      3\,\left( {\hat{U}_\phi}^2 - {\hat{U}_\psi}^2 + \frac{\gamma}{2} \,\hat{\Omega}\right)
          \right) },\\
& \hat{U}_\phi'=3\,m\,\hat{W}^2 + 3\,Q_0\,\hat{U}_\phi\,\left( -1 + {\hat{U}_\phi}^2 - {\hat{U}_\psi}^2 + 
     \frac{\gamma}{2}\,\hat{\Omega} \right), \\
& \hat{U}_\psi'=-3\,n\,\hat{W}^2 + 3\,Q_0\,\hat{U}_\psi\,\left( -1 + {\hat{U}_\phi}^2 - {\hat{U}_\psi}^2 + 
     \frac{\gamma}{2}\,\hat{\Omega} \right),\\
&  \hat{W}'=-3\,\hat{W}\left(m\,\hat{U}_\phi+n\,\hat{U}_\psi-Q_0\,\left( {\hat{U}_\phi}^2 - {\hat{U}_\psi}^2 + 
     \frac{\gamma}{2}\,\hat{\Omega} \right)\right).\label{ds2}              
\end{align}
}

There is also an auxiliary evolution equation 

\begin{equation}
\hat{\Omega}'=-Q_0\,\left( -2\,\left( {\hat{U}_\phi}^2 - {\hat{U}_\psi}^2 \right)  + \gamma \,\left( 1 - \hat{\Omega}  \right)  \right) \, \hat{\Omega}. \label{eqOmega}
\end{equation}

It is useful to express some cosmological parameters in terms of our state variables. 

$$\left(\Omega_m,\Omega_{de},\Omega_k\right)=\left(\hat{\Omega},1-\hat{\Omega},Q_0^2-1\right)/Q_0^2.$$

Observe that the system (\ref{ds2}, \ref{eqOmega}) is invariant under the transformation of coordinates

\begin{equation}
\left(\hat{\tau}, Q_0, \hat{U}_\phi, \hat{U}_\psi,\hat{W},\hat{\Omega}\right)\rightarrow \left(-\hat{\tau}, -Q_0, -\hat{U}_\phi, -\hat{U}_\psi, \hat{W},\hat{\Omega}\right)\label{discrete2}.
\end{equation}
Thus, it is sufficient to discuss the behavior in one part of the phase space. The dynamics in the other part being obtained via the transformation (\ref{discrete2}).

\begin{rem}
The function 

\begin{equation}
N=\frac{\left(n\,\hat{U}_\phi+m\,\hat{U}_\psi\right)^2\,\hat{\Omega}^2}{\left(1-Q_0^2\right)^{3}},\; N'=-6\gamma\,Q_0\,N\label{monotonic2}
\end{equation} is monotonic in the regions $Q_0<0$ and $Q_0>0$ for $Q_0^2\neq 1,\, n\,\hat{U}_\phi+n\,\hat{U}_\psi\neq 0,\, \hat{\Omega}>0.$ Hence, there can be no periodic orbits or recurrent orbits in the interior of the phase space. Furthermore, it is possible to obtain global results. From the expression $N$ we can immediately see that asymptotically $Q_0^2\rightarrow 1$ or $n \hat{U}_\phi+m\tilde{U}_\psi\rightarrow 0$ or $\hat{\Omega}\rightarrow 0.$
\end{rem}

In the table \ref{T4} it is summarize the location, existence conditions and the eigenvalues of the linearized system around each equilibrium point. In the following we will characterize the dynamical behavior of the cosmological solutions associated with them.

The equilibrium points ${}_\pm\hat{K},{}_\pm\hat{F}, {}_\pm \hat{SF}$ and  ${}_\pm \hat{MS}$ represents flat FRW solutions. They corresponds to the same cosmological solutions as the not hatted ones. We submit the reader to the early sections for the physical interpretation of them. Although ${}_\pm \hat{CS}$  represent a curvature scaling solution, it is different from ${}_\pm {CS}$ in the sense that it represents a positive curvature model.  

By the above discussion, we will focus here on the dynamical character of the equilibrium points, not in its physical interpretation. As before, due the symmetry (\ref{discrete2}), we will characterize the equilibrium points corresponding to the ``positive'' branch. The dynamical behavior of the equilibrium points in the ``negative'' branch is determined by the transformation (\ref{discrete2}).  

Observe that the equilibrium points ${}_\pm \hat{MS}$ and  ${}_\pm \hat{SF}$ coincide as $\delta\rightarrow \frac{\gamma}{2}^+$ and ${}_\pm \hat{CS}$ and  ${}_\pm \hat{SF}$ coincide as $\delta\rightarrow\frac{1}{3}^-.$ Additionally, ${}_+ \hat{SF}$ (resp., ${}_- \hat{SF}$) coincides with a point in the arc ${}_+\hat{K}_+$ (resp., ${}_-\hat{K}_-$) as $\delta\rightarrow 1^-.$ These values of $\delta$ where equilibrium points coincide corresponds to bifurcations. 
Combining the above linear analysis, with the information we have using monotonic functions we have the following
\begin{prop}
The past and future attractors for the quintom model with $k=1$ are as follows:

\begin{enumerate}
\item For 
$Q_0=-1$ 
\begin{enumerate}
\item The past attractors are:
 \begin{enumerate}
\item ${}_-\hat{SF}$ if $0\leq\gamma\leq 2,\,\delta<\sfrac{\gamma}{2}$.
\item ${}_-\hat{MS}$ if $0<\gamma<2,\,\delta>\sfrac{\gamma}{2}$.  
\end{enumerate}

\item The future attractor is ${}_-\hat{K}_\pm$ if $n U_\psi^\star\pm m\sqrt{1+{U_\psi^\star}^2}>-1$.
\end{enumerate}

\item 
For $-1<Q_0<0$ 
\begin{enumerate}
\item The past attractor are:

\begin{enumerate} 
\item ${}_-\hat{SF}$ if $\frac{2}{3}\leq\gamma\leq 2,\,\delta<\sfrac{1}{3}$ or $0\leq\gamma\leq \sfrac{2}{3},\,\delta<\sfrac{\gamma}{2}$.
\item ${}_-\hat{MS}$ if $0<\gamma<\sfrac{2}{3},\,\delta>\sfrac{\gamma}{2}$.
\end{enumerate}

\item The future attractor is ${}_-\hat{K}_\pm$ if $n U_\psi^\star\pm m\sqrt{1+{U_\psi^\star}^2}>-1$.

\end{enumerate}

\item 
For 
$0<Q_0<1$ 
\begin{enumerate}
\item The past attractor is
${}_+\hat{K}_\pm\,\text{if}\;n U_\psi^\star\pm m\sqrt{1+{U_\psi^\star}^2}<1$. 
\item The future attractor is 
\begin{enumerate} 
\item ${}_+\hat{SF}$ if $\frac{2}{3}\leq\gamma\leq 2,\,\delta<\sfrac{1}{3}$ \\ or $0\leq\gamma\leq \sfrac{2}{3},\, \delta<\sfrac{\gamma}{2}$.
\item 
${}_+\hat{MS}$ if $0<\gamma<\sfrac{2}{3},\,\delta>\sfrac{\gamma}{2}$.
\end{enumerate}
\end{enumerate}

\item 
For 
$Q_0=1$ 
\begin{enumerate}
\item 
The past attractor is
${}_+\hat{K}_\pm$ if $n U_\psi^\star\pm m\sqrt{1+{U_\psi^\star}^2}<1$.
\item The future attractors are: 
\begin{enumerate} 
\item ${}_+\hat{SF}$  if $0\leq\gamma\leq 2,\,\delta<\sfrac{\gamma}{2}$.
\item ${}_+\hat{MS}$ if $0<\gamma<2,\,\delta>\sfrac{\gamma}{2}$.
\end{enumerate}
\end{enumerate}
\end{enumerate}
\end{prop}
\begin{table*}[t!]
\begin{tabular}{|c|c|c|c|}
\hline
Label & $(Q_0, \hat{U}_\phi,\hat{U}_\psi,\hat{W})$& Existence& Eigenvalues\\
\hline
$S_\pm$ & $(0,{\hat{U}_\phi},{\hat{U}_\psi},0),$ ${\hat{U}_\phi}^2-{\hat{U}_\psi}^2=\frac{2-3\gamma}{3(2-\gamma)}$ & $0\leq\gamma\leq \sfrac{2}{3}$ & $0,\pm \sqrt{2(2-3\gamma)},-3n\tilde{U}_\psi-3 m\hat{U}_\phi$ \\[0.2cm]
${}_{\pm}\hat{K}_{\pm}$ & $(\epsilon,\pm\sqrt{1+{U_\psi^\star}^2},U_\psi^\star,0)$ & All $m$ and $n$ & $4\epsilon,0,3\left(\epsilon-n U_\psi-m U_\phi\right),3(2-\gamma)\epsilon$ \\[0.2cm]
${}_{\pm}\hat{F}$ & $(\epsilon,0,0,0)$ & All $m$ and $n$ & $\left(3\gamma-2\right)\epsilon,\frac{3}{2}\gamma\epsilon,3\left(\frac{\gamma}{2}-1\right)\epsilon[\times 2]$ \\[0.2cm]
${}_\pm \hat{SF}$ & $(\epsilon,m\epsilon,-n\epsilon,\sqrt{1-\delta})$ & $\delta<1$ & $-2\left(1-3\delta\right)\epsilon,-3\left(1-\delta\right)\epsilon[\times 2],3\left(2\delta-\gamma\right)\epsilon$ \\[0.2cm]
${}_\pm \hat{CS}$  & $(\sqrt{3\delta}\epsilon,\frac{m \epsilon}{\sqrt{3\delta}},-\frac{n \epsilon}{\sqrt{3\delta}},\sqrt{\frac{2}{3}})$ & $0<\delta<\frac{1}{3}$& $-\sqrt{3\delta}\epsilon\pm\sqrt{4-9\delta},-2\sqrt{3\delta}\epsilon,(2-3\gamma)\sqrt{3\delta}\epsilon$ \\[0.2cm]
${}_{\pm}\hat{MS}$ & $(\epsilon,\frac{m\gamma}{2\delta},-\frac{n\gamma}{2\delta},\frac{\sqrt{(2-\gamma)\gamma}}{2\sqrt{\delta}},\sqrt{1-\frac{\gamma}{2\delta}})$ & $\delta>\frac{\gamma}{2}$  & $-\left(2-3\gamma\right)\epsilon,3\left(-1+\frac{\gamma}{2}\right)\epsilon, 3\lambda^\pm\epsilon$ \\[0.4cm]
\hline
\end{tabular}
\caption{\label{T4} equilibrium points in the quintom model with $k=+1.$ We use the same notation as in table \ref{T1}. If $\gamma =\frac{2}{3},$ the set of equilibrium points, $S,$ (corresponding to static solutions) reduces to two sets of equilibrium points analogous to $V_{1,2}$ (see table \ref{T1}). When the flow is restricted to the invariant sets $Q_0=\pm 1,$ the eigenvalues associated to the equilibrium points ${}_{\pm}\hat{F},$ ${}_\pm \hat{SF}$ and ${}_\pm \hat{MS}$ an to the equilibrium sets ${}_{\pm}\hat{K}_{\pm},$ are, in each case, the same as those displayed, but the first from the left.}
\end{table*}

\section{Integrability and algebraic solution}
\label{integrability}

In this section we examine the integrability for the gravitational field
equations of the quintom model by using two methods for the study of the
integrability of dynamical systems. In particular we apply the symmetry
analysis, we determine point transformations which leaves invariants the
differential equations which provide also conservation laws. The second
method that we assume is the singularity analysis where we are able to write
the algebraic solution for the cosmological quintom model of our
consideration.

\subsection{Symmetry analysis}
\label{symm}
The symmetry analysis has an important role for the study of the
integrability and the determination of analytical solutions in gravitational
theories \cite{Maharaj1,Maharaj2,Maharaj3}. There is a plethora of
applications in the literature of the symmetry analysis in cosmological
studies, in the dark energy models and in the modified theories of gravity,
for instance see \cite{ns01,ns02,ns03,ns04,ns05} and references therein.

Consider now the point-like Lagrangian for the quintom cosmological model (%
\ref{expquintomlag}) with potential (\ref{ExponentialPot}) in the spatially
flat geometry and without any other matter source%
\begin{align}
& L\left( a,\dot{a},\phi ,\dot{\phi},\psi ,\dot{\psi}\right)= \nonumber \\ & -3a\dot{a}%
^{2}+\frac{1}{2}a^{3}\left( \dot{\phi}^{2}-\dot{\psi}^{2}\right)
-V_{0}a^{3}e^{-\sqrt{6}(m\phi +n\psi )}.  \label{lan.01}
\end{align}

The field equations are the Euler-Lagrange equations of the later point-like
Lagrangian plus the Hamiltonian constraint
\begin{equation}
\mathcal{H}\equiv -3a\dot{a}^{2}+\frac{1}{2}a^{3}\left( \dot{\phi}^{2}-\dot{%
\psi}^{2}\right) -V_{0}a^{3}e^{-\sqrt{6}(m\phi +n\psi )}=0.
\label{lan.02}
\end{equation}%
which is the first Friedmann equation.

By following the Noether symmetry analysis, for a recent review on the
approach and more details for the method we refer the reader to \cite{ns04},
we find that the Lagrangian (\ref{lan.01}) admits the following two Noether
symmetries, except the trivial autonomous symmetry,%
\begin{equation}
X_{1}=n\partial _{\phi }-m\partial _{\psi }~,~X_{2}=\frac{2}{3}a\partial
_{a}+\frac{2\sqrt{6}}{3\left( m+n\right) }\left( \partial _{\phi }+\partial
_{\psi }\right) ,  \label{lan.03}
\end{equation}%
where the corresponding Noetherian conservation laws
\begin{align}
&\Phi _{1}=a^{3}\left( n\dot{\phi}+m\dot{\psi}\right) ,  \label{lan.04}
\\
&\Phi _{2}=-4a^{2}\dot{a}+\frac{2\sqrt{6}}{3\left( m+n\right) }a^{3}\left(
\dot{\phi}-\dot{\psi}\right) .  \label{lan.05}
\end{align}

The three conservation laws, $\mathcal{H}$,~$\ \Phi _{1}$ and $\Phi _{2}$
are linear independent, and in-involution, that is $\left\{ \Phi _{1},\Phi
_{2}\right\} =0,~\left\{ \Phi _{1},\mathcal{H}\right\} =0$ and $\left\{ \Phi
_{2},\mathcal{H}\right\} =0$, where $\left\{ ,\right\} $ denotes the Poisson
bracket. Thus, we can refer that the gravitational field equations described
by the Lagrangian (\ref{lan.02}) form a Liouville integrable dynamical
system \cite{arnold}.

Notice that the above conservation laws can be written in terms of the phase plane variables as 
\begin{align}
&\frac{\Phi_1}{\sqrt{6} \dot a a}=n x_\phi +m x_\psi,\\
&\frac{\Phi_2}{4 \dot a a}=\frac{x_\phi-x_\psi}{m+n}-1,
\end{align}
such that at the invariant set $\{{\bf x}\in \mathbb{R}^2: n x_\phi +m x_\psi=0\} $ we have the trivial charge $\Phi_1=0$, whereas in the invariant set 
$\{{\bf x}\in \mathbb{R}^2:\frac{x_\phi-x_\psi}{m+n}=1\}$  we have the trivial charge $\Phi_2=0$.

Because of the nonlinearity of the differential equations it is not possible
to write the analytical solution in closed-form by using well-known
functions. Hence, we continue our analysis with the singularity analysis
where the solution can be written by using Laurent expansions.

\subsection{Singularity analysis\newline
}

Except from the symmetry analysis, singularity analysis is another approach
to study the integrability of a given dynamical system. In singularity
analysis integrability is not concerned with the display of explicit
functions but with the demonstration of \ a specific property, such that:
the existence of a Laurent series for each of the dependent variables. The
series may not be summable to an explicit form, but does represent an
analytic function apart from any singular points. Moreover, the essential
feature of this Laurent series is that it is an expansion about a particular
type of movable critical point, a pole.

While the symmetry analysis is related with the existence of closed-form
solutions or solutions expressed in terms of well-known first-order
differential equations. There is a choice of the type of definitions for the
symmetries which can be applied in the symmetry analysis, where different
conclusions can be occurs. On the other hand, singularity analysis is
straightforward in its application as it does not offer so many choices.
Both methods are complementary \cite{leachrev}.

The main steps for the application of the singularity analysis for a given
differential equation are described by the ARS algorithm (from the initial
of Ablowitz, Ramani and Segur) \cite{Abl1,Abl2,Abl3}. The latter algorithm
has been used in various cosmological models to determine the integrability
and write algebraic solutions, for instance see \cite%
{sig1,cots,sig2,sig3,sig4,sig5} and references therein. For a demonstration
on the application of the ARS algorithm we refer the reader in the review of
Ramani et al. \cite{Ramani89}.

Consider the dynamical system (\ref{eqxphi}), (\ref{eqxvphi}) and (\ref{eqy}%
) which describes the evolution of the universe in a spatially flat FRW
universe. We search for singular behaviour of the form%
\begin{eqnarray}
x_{\phi }\left( t\right)  &=&x_{\phi 0}\left( t-t_{0}\right) ^{p_{1}}~\ ,~
\label{sg.01} \\
x_{\psi }\left( t\right)  &=&x_{\psi 0}\left( t-t_{0}\right)
^{p_{2}}~,~ \\
y\left( t\right)  &=&y_{0}\left( t-t_{0}\right) ^{p_{3}}
\end{eqnarray}%
where $t_{0}$ is an arbitrary constant and denotes the position of the
singularity. Thus we find the unique leading-order behavior to be%
\begin{eqnarray}
x_{\phi }\left( t\right)  &=&x_{\phi 0}\left( t-t_{0}\right) ^{-\frac{1}{2}%
}\ ,~  \label{sg.02} \\
x_{\psi }\left( t\right)  &=&x_{\psi 0}\left( t-t_{0}\right) ^{-\frac{1%
}{2}}~,~ \\
y\left( t\right)  &=&y_{0}\left( t-t_{0}\right) ^{-\frac{1}{2}}
\end{eqnarray}%
with constraint equation%
\begin{equation}
1-\left( x_{\phi 0}\right) ^{2}+\left( x_{\psi _{0}}\right) ^{2}-\left(
y_{0}\right) ^{2}=0.  \label{sg.03}
\end{equation}%
From the latter it follows that two of the coefficients of the leading-order
terms are arbitrary constants. In particular are the two constants of
integrations, recall that $t_{0}$ is the third integration constant. Hence,
we can conclude that equations (\ref{eqxphi}), (\ref{eqxvphi}) and (\ref{eqy}%
) form an integrable dynamical system. However, in order to verify it we
proceed with the next steps of the ARS algorithm.

In order to determine the resonances, we substitute in the dynamical system %
\ref{eqxphi}), (\ref{eqxvphi}) and (\ref{eqy}) \ the following expressions%
\begin{eqnarray}
x_{\phi }\left( t\right)  &=&x_{\phi 0}\left( t-t_{0}\right) ^{-\frac{1}{2}%
}+\varepsilon _{1}\left( t-t_{0}\right) ^{-\frac{1}{2}+s}\  \\
~x_{\psi }\left( t\right)  &=&x_{\psi 0}\left( t-t_{0}\right) ^{-\frac{%
1}{2}}+\varepsilon _{2}\left( t-t_{0}\right) ^{-\frac{1}{2}+s}~ \\
~y\left( t\right)  &=&y_{0}\left( t-t_{0}\right) ^{-\frac{1}{2}}+\varepsilon
_{3}\left( t-t_{0}\right) ^{-\frac{1}{2}+s}
\end{eqnarray}%
in which $\varepsilon _{1},~\varepsilon _{2},~\varepsilon _{3}$ are
infinitesimal parameters.\ We follow the steps described in \cite{Ramani89}
for the determination of the resonances in systems of differential equations
we find the three resonances to be%
\begin{equation*}
s_{1}=-1~,~s_{2}=0~\text{and }s_{3}=0.
\end{equation*}

Resonance $s_{1}=-1$ is important to exist in order the singularity analysis
to succeed. In particular it is related with the movable singularity. On the
other hand, the values zero of the other two resonances tell us that two of
the coefficient terms for the leading-order behavior are integration
constants of the problem. Moreover, because $s_{2},~s_{3}= 0$, the
solution of the dynamical system is expressed by Right Painlev\'{e} series,
that is,%
\begin{equation}
x_{\phi }\left( t\right) =x_{\phi 0}\left( t-t_{0}\right) ^{-\frac{1}{2}%
}+x_{\phi _{1}}\left( t-t_{0}\right) +\sum\limits_{i=2}x_{\phi _{i}}\left(
t-t_{0}\right) ^{-\frac{1}{2}+\frac{i}{2}},  \label{sg.04}
\end{equation}%
\begin{equation}
x_{\psi }\left( t\right) =x_{\psi 0}\left( t-t_{0}\right) ^{-\frac{1}{2%
}}+x_{\psi _{1}}\left( t-t_{0}\right) +\sum\limits_{j=2}x_{\psi
_{j}}\left( t-t_{0}\right) ^{-\frac{1}{2}+\frac{j}{2}},  \label{sg.05}
\end{equation}%
\begin{equation}
y\left( t\right) =y_{0}\left( t-t_{0}\right) ^{-\frac{1}{2}}+y_{1}\left(
t-t_{0}\right) +\sum\limits_{k=2}y_{k}\left( t-t_{0}\right) ^{-\frac{1}{2}+%
\frac{k}{2}},  \label{sg.06}
\end{equation}%
where the first coefficient constants $x_{\phi _{1}},~x_{\psi _{2}}$ and $%
y_{1}$ are calculated to be
\begin{eqnarray}
x_{\phi _{1}} &=&\frac{2}{3}m\left( 3+4\left( x_{\phi _{0}}\right)
^{2}\right) \left( y_{0}\right) ~^{2},~ \\
x_{\psi _{1}} &=&-\frac{2}{3}\left( 3n-4mx_{\phi _{0}}x_{\psi
_{0}}\right) \left( y_{0}\right) ^{2}~,~
\end{eqnarray}%
\begin{equation}
y_{1}=\frac{y_{0}}{3}\left( mx_{\phi _{0}}\left( \left( y_{0}\right)
^{2}-1\right) -3nx_{\psi _{0}}\right) .
\end{equation}

Thus we can conclude that the dynamical system (\ref{eqxphi}), (\ref{eqxvphi}%
) and (\ref{eqy}) passes the singularity test and its algebraic solution is
given by the Right Painlev\'{e} Series (\ref{sg.04}), (\ref{sg.05}), (\ref%
{sg.05}). From the latter we can infer that the leading-order term (\ref%
{sg.02}), i.e. the dominant behavior close to the singularity is an
unstable solution.

We perform the same analysis for the complete dynamical system (\ref{0eqxphi}%
), (\ref{0eqxvphi}), (\ref{0eqy}) and (\ref{0eqz}), where we find similar
results. In particular the leading-order behavior is%
\begin{eqnarray}
x_{\phi }\left( t\right)  &=&x_{\phi 0}\left( t-t_{0}\right) ^{-\frac{1}{2}%
}\ ,~x_{\psi }\left( t\right) =x_{\psi 0}\left( t-t_{0}\right) ^{-%
\frac{1}{2}}~,  \label{sg.07} \\
~y\left( t\right)  &=&y_{0}\left( t-t_{0}\right) ^{-\frac{1}{2}}~,~z\left(
t\right) =z_{0}\left( t-t_{0}\right) ^{-\frac{1}{2}}
\end{eqnarray}%
with constraint equation%
\begin{equation}
1-\left( \left( x_{\phi 0}\right) ^{2}-\left( x_{\psi _{0}}\right)
^{2}\right) -\left( y_{0}\right) ^{2}+\frac{\left( z_{0}\right) ^{2}}{3}%
\left( 3\gamma -2\right) =0.  \label{sg.08}
\end{equation}%
and resonances $s_{1}=-1$, $s_{2}=0$, $s_{3}=0$ and $s_{4}=0$. Hence, we
have the following proposition.

\begin{prop}
The dynamical system described the four first-order equations (\ref{0eqxphi}%
), (\ref{0eqxvphi}), (\ref{0eqy}) and (\ref{0eqz}), passes the singularity
test with leading-order terms given by (\ref{sg.07}). The algebraic solution
is given by the Right Painlev\'{e} Series%
\begin{equation}
\mathbf{x}\left( t\right) =\mathbf{x}_{0}\left( t-t_{0}\right) ^{-\frac{1}{2}%
}+\mathbf{x}_{1}\left( t-t_{0}\right) +\sum\limits_{i=2}\mathbf{x}%
_{i}\left( t-t_{0}\right) ^{-\frac{1}{2}+\frac{i}{2}},  \label{sg.09}
\end{equation}%
where $\mathbf{x=}\left( x_{\phi },x_{\psi },y,z\right) $ and constraint
equation (\ref{sg.08}). From the form of the Laurent expansions we refer
that the leading-order term describe a past attractor for the dynamical
system.
\end{prop}

\section{Concluding Remarks}\label{conclusions}

We discussed some results concerning the asymptotic dynamics of
quintom cosmologies with exponential potential. Using in a combined way several tools of the dynamical systems theory such as the linear stability analysis, Normal Forms calculations,  Center manifold and Invariant Manifold Theorems, by developing Monotonic functions and implementing numerical solutions, one can find information bout the early time and late time behavior and at the intermediate stages of the evolution.

We have divided the analysis of the negative curvature and zero curvature models from the analysis of positive curvature models. 

For negative or zero curvature models the
physical behavior of a typical quintom cosmology as as follow. For ever expanding cosmologies, near the big-bang a typical model behaves like a flat FL model with stiff fluid (i.e. the dark energy mimics a stiff fluid) represented by the equilibrium set ${}_+K_+$ or by ${}_+K_-$ (parameterized by the real value $U_\psi^\star$) depending if $n U_\psi^\star+ m\sqrt{1+{U_\psi^\star}^2}$ or $n U_\psi^\star- m\sqrt{1+{U_\psi^\star}^2}$ is less than $1.$ If $\sfrac{2}{3}<\gamma<2$ and $\delta<0$ the late time dynamics is determined by ${}_+SF$, i.e. the model is accelerating, close to flatness ($\Omega_k\rightarrow 0$) and dominated by phantom dark energy ($\Omega_{de}\rightarrow 1$). This means, that typically, ever expanding quintom model crosses the phantom divide (it EoS parameter takes values less than $-1$). The intermediate dynamics is will be governed to a large extent by the fixed points ${}_+CS,$ ${}_+MS,$ and ${}_+M,$ which have a lower dimensional stable manifold. For $H$ large and positive (i.e., in the invariant set $D=1$), the late time dynamics is determined by the equilibrium point ${}_+SF$ provided $\delta<\sfrac{1}{3}$ which represents quintessence ($-1 < q < 0$, i.e. $-1 < w < -\sfrac{1}{3}$ ) or a phantom field ($q < -1$, i.e. $w < -1$) if $\delta> 0$ or $\delta< 0$ respectively. If $\delta=0$ it represents a de Sitter cosmological model. Curvature dominates the late time evolution in the invariant set $D=1$ whenever $\sfrac{2}{3}<\gamma<2$ and $\delta>\sfrac{1}{3}.$ 

On the other hand, for contracting models, the typical behavior is, in one sense, the reverse of the above, that is, if $\sfrac{2}{3}<\gamma<2$ and $\delta<0$ the early time dynamics is determined by ${}_-SF$, i.e. the model is accelerating, close to flatness ($\Omega_k\rightarrow 0$) and dominated by phantom dark energy ($\Omega_{de}\rightarrow 1$). The intermediate dynamics is will be governed to a large extent by the fixed points ${}_-CS,$ ${}_-MS,$ and ${}_-F,$ which have a lower dimensional stable manifold. For $H$ large and negative (i.e., in the invariant set $D=-1$), the early time dynamics is determined by the equilibrium point ${}_+SF$ provided $\delta<\sfrac{1}{3}.$ Curvature dominates the early time evolution in the invariant set $D=1$ whenever $\sfrac{2}{3}<\gamma<2$ and $\delta>\sfrac{1}{3}.$ A typical model behaves at late times like a flat FL model with stiff fluid (i.e. the dark energy mimics a stiff fluid) represented by the equilibrium set ${}_-K_+$ or by ${}_-K_-$ (parameterized by the real value $U_\psi^\star$) depending if $n U_\psi^\star+ m\sqrt{1+{U_\psi^\star}^2}$ or $n U_\psi^\star- m\sqrt{1+{U_\psi^\star}^2}$ is greater than $-1.$

The physical interpretation of the equilibrium points corresponding to positive curvature models was given in \cite{Leon:2009ce}. The physical behavior of a typical closed quintom cosmology is as follows \cite{Leon:2009ce}. For ever expanding cosmologies, near the big-bang a typical model behaves like
a flat FRW model with stiff fluid represented by the equilibrium set ${}_{+}\hat{K}_{+}$ or by
${}_{+}\hat{K}_{-}$, depending on the choice of the values of the free parameters $m, n$ and
$x_\psi^{\star}$. If $\delta < 1/3$ 
and $0 < Q_0 < 1$ the late time dynamics is determined by ${}_{+} \hat{SF}$,
(with the same physical properties as ${}_{+} {SF}$). The intermediate dynamics will
be governed to a large extent by the fixed points ${}_{+}\hat{CS}$, ${}_{+}\hat{MS}$ and ${}_{+}\hat{F}$ which
have the highest lower-dimensional stable manifold. For flat models (i.e., in the
invariant set $Q_0 = 1$), the late time dynamics is determined by the equilibrium point
${}_{+} \hat{SF}$ provided $\delta<1/2$ or ${}_{+} \hat{MS}$ provided $\delta>1/2$
.

For contracting models, the typical behavior is, in one sense, the reverse of
the above. If $\delta<1/3$ and $-1<Q_0<0$ the early time dynamics is determined by
${}_{-} \hat{SF}$. The intermediate dynamics will be governed to a large extent by the fixed
points ${}_{-} \hat{CS}$, ${}_{-} \hat{MS}$, ${}_{-} \hat{F}$ which have the highest lower-dimensional stable
manifold. For flat models (i.e., in the invariant set $Q_0=-1$), the early time
dynamics is determined by the equilibrium point ${}_{-} \hat{SF}$ (or ${}_{-} \hat{MF}$ ) provided $\delta<1/2$ ($\delta>1/2$). A typical model behaves at late times like a flat FRW model with stiff
fluid (i.e. the dark energy mimics a stiff fluid) represented by the equilibrium set
${}_{-} \hat{K}_{+}$ or by ${}_{-} \hat{K}_{-}$ depending on the choice of the values of the free parameters
$m, n$ and $x_\psi^{\star}$. 

Furthermore, we applied the symmetry analysis and the singularity analysis to determine the integrability of the quintom model. By using the Noether point symmetries in the case of the flat FRW geometry without any matter source we found that the point-like Lagrangian admits three linear-independent conservations laws which are in involution. Therefore, the field equations form a Liouville integrable dynamical system. However, because of the nonlinearity of the field equations that analytical solution can not be written in close-form expression. 

On the other hand, by using the singularity analysis we were able to prove the integrability either when matter source exists and also the spatial curvature is not zero. We wrote the analytic solution of the field equations by using Painlev\'{e} Series by starting from a unstable singular solution. 

The two methods, symmetries and singularities, are complementary, while in a future analysis will be of special interest to determine new analytical solutions for other kind of potentials. 

\vskip .1in \noindent {\bf {Acknowledgments}}
GL thanks to Department of Mathematics and to Vicerrector\'ia de Investigaci\'on y Desarrollo
Tecnol\'ogico at Universidad Cat\'olica del Norte for financial support. AP acknowledges financial support of FONDECYT grant No. 3160121.
JLM acknowledges to Consejo Nacional de Ciencia y Tecnología (CONACYT) of Mexico for  financial support throughout the Phd program on Water Sciences and Technology, Grant No. 003497.

\appendix

\section{Invariant sets and monotonic functions}
\label{monotonic-th}

In the following we shall investigate some invariant sets that can be trivially identified and we construct properly monotonic functions defined on them.

Recall the definition and results: 
\begin{defn}[Invariant set]
Let $S\subset\mathbb{R}^n$ be a set. $S$ is called an invariant
set under the vector field \begin{equation}
\mathbf{x}'={\bf X}(\mathbf{x})\label{T1.2}
\end{equation} if $\mathbf{y}\in S$
implies $\mathbf{x}(\tau, \mathbf{y})\in S$ (where $\mathbf{x}(0,
\mathbf{y})=\mathbf{y}$) for all $\tau\in\mathbb{R}.$ If we
consider the property valid for $\tau\geq 0$ we say that $S$ is
positively invariant. On the other hand, if the property is valid
for $\tau\leq 0$ we say that $S$ is negatively invariant.
\end{defn}

\begin{prop}[Proposition 4.1, \cite{reza} p. 92]\label{Proposition 4.1} Consider the
autonomous vector field \eqref{T1.2} with flow $\mathbf{g}^\tau.$
Let be defined a $C^1$ function $Z:\mathbb{R}^n\rightarrow
\mathbb{R}$ which satisfies $Z'\equiv \nabla Z\cdot {\bf
X}(\mathbf{x})=\alpha Z$ where
$\alpha:\mathbb{R}^n\rightarrow\mathbb{R}$ is a continuous
function. Then, the subsets of $\mathbb{R}^n$ defined by
$\left\{\mathbf{x}\in\mathbb{R}^n| Z(\mathbf{x})\lesseqqgtr
0\right\}$ are invariant sets for the flow $\mathbf{g}^\tau.$
\end{prop}

\begin{defn}[Definition 4.8 \cite{reza}, p. 93]\label{Definition 4.8} Let $\mathbf{g}^\tau(\mathbf{x})$ be a flow on $\mathbb{R}^n,$ let $S$ be an invariant set of $\mathbf{g}^\tau(\mathbf{x})$ and let $Z:S\rightarrow\mathbb{R}$ be a continuous function. $Z$
is monotonic decreasing (increasing) function for the flow
$\mathbf{g}^\tau(\mathbf{x})$ means that for all $\mathbf{x}\in
S,$ $Z(\mathbf{g}^\tau(\mathbf{x}))$ is a monotonic decreasing
(increasing) function of $\tau.$

\end{defn}

The existence of a continuous monotone function on an invariant set $S$ simplifies the orbits in $S$ significantly. As states the following proposition.

\begin{prop}[Proposition 4.2 (\cite{reza}, pp 93)]\label{Proposition 4.2}  The existence of a continuous monotone function on an invariant set $S$ implies that $S$ contains no equilibrium points, periodic orbits, recurrent orbits or homoclinic orbits.
\end{prop}

\begin{thm}[Monotonicity Principle ({\bf{Proposition A.I}} in \cite{LeBlanc:1994qm} developed in collaboration with M Glaum, cited as {\bf Theorem 4.12} in \cite{reza})]\label{theorem 4.12}

Let $\mathbf{g}^\tau(\mathbf{x})$ be a flow on $\mathbb{R}^n$ with
$S$ an invariant set. Let $Z: S\rightarrow\mathbb{R}$ be a
$C^1\left(\mathbb{R}^n\right)$ function whose range is the
interval $(a,\;b)$ where $a\in \mathbb{R} \cup \{-\infty\},$ $b\in
\mathbb{R} \cup \{+\infty\},$ and $a<b.$ If $Z$ is decreasing on
orbits in $S,$ then for all $\mathbf{x}\in S$,
$\omega(\mathbf{x})\subset\{\mathbf{s}\in \bar S\setminus
S|lim_{\mathbf{y}\rightarrow \mathbf{s}} Z(\mathbf{y})\neq b \}$
and $\alpha(\mathbf{x})\subset\{\mathbf{s}\in \bar S\setminus
S|lim_{\mathbf{y}\rightarrow \mathbf{s}} Z(\mathbf{y})\neq a \}.$
\end{thm}

\section{Normal Forms for vector fields}\label{NFTheory}

Let ${\bf X}:\mathbb{R}^n\rightarrow \mathbb{R}^n$ be a smooth
vector field satisfying ${\bf X}({\bf 0})={\bf 0}.$ We can
formally construct the Taylor expansion of ${\bf x}$ about ${\bf
0},$ namely, ${\bf X}={\bf X}_1+{\bf X}_2+\ldots +{\bf
X}_k+{O}(|{\bf x}|^{k+1}),$ where ${\bf X}_r\in H^r,$ the real
vector space of vector fields whose components are homogeneous
polynomials of degree $r$. For $r=1$ to $k$ we write

\ben
&{\bf X}_r({\bf x})=\sum_{m_1=1}^{r}\ldots\sum_{m_n=1}^{r}\sum_{j=1}^{n}{\bf X}_{{\bf m},j}{{\bf x}}^{{\bf m}}{\bf e}_j,\nonumber\\
& \sum_i m_i=r, \een

Observe that ${\bf X}_1={\bf DX(\mathbf{0})}{\bf x}\equiv {\bf
A}{\bf x},$ i.e., the matrix of derivatives. 

We have the following result
\begin{thm}[theorem 2.3.1 in \cite{arrowsmith}]\label{NFTheorem}
Given a smooth vector field $\bf X({\bf x})$ on $\mathbb{R}^n$
with ${\bf X(0)=0},$ there is a polynomial transformation to new
coordinates, ${\bf y},$ such that the differential equation ${\bf
x}'={\bf X}({\bf x})$ takes the form ${\bf y}'={\bf J}{\bf
y}+\sum_{r=1}^N {\bf w}_r({\bf y})+{O}(|{\bf y}|^{N+1}),$ where
${\bf J}$ is the real Jordan form of ${\bf A}={\bf D X}({\bf 0})$
and ${\bf w}_r\in G^r,$ a complementary subspace of $H^r$ on
$B^r={\bf L_A}(H^r),$ where ${\bf
L_A}$ is the linear operator that assigns to ${\bf h(y)}\in H^r$
the Lie bracket of the vector fields ${\bf A y}$ and ${\bf h(y)}$:
\ben {\bf L_A}: H^r& &\rightarrow H^r\nonumber\\
     {\bf h}  & & \rightarrow  {\bf L_A} {\bf h (y)}={\bf D h(y)} {\bf A y}- {\bf A h(y)}.
\een
\end{thm}

\end{document}